\documentclass[%
preprint,
showpacs,
 amsmath,amssymb,
 aps,
]{revtex4-1}

\usepackage{graphicx}
\usepackage{dcolumn}
\usepackage{bm}

\begin{document}


\title{Nuclear Lattice Simulations using Symmetry-Sign Extrapolation}

\author{Timo~A.~L\"{a}hde}
\affiliation{Institute~for~Advanced~Simulation, Institut~f\"{u}r~Kernphysik, and
J\"{u}lich~Center~for~Hadron~Physics,~Forschungszentrum~J\"{u}lich,
D-52425~J\"{u}lich, Germany}

\author{Thomas Luu}
\affiliation{Institute~for~Advanced~Simulation, Institut~f\"{u}r~Kernphysik, and
J\"{u}lich~Center~for~Hadron~Physics,~Forschungszentrum~J\"{u}lich,
D-52425~J\"{u}lich, Germany}

\author{Dean~Lee}
\affiliation{Department~of~Physics, North~Carolina~State~University, Raleigh, 
NC~27695, USA}

\author{Ulf-G.~Mei{\ss }ner}
\affiliation{ Helmholtz-Institut f\"ur Strahlen- und Kernphysik and Bethe Center for Theoretical Physics, \\
Universit\"at Bonn,  D-53115 Bonn, Germany}
\affiliation{Institute~for~Advanced~Simulation, Institut~f\"{u}r~Kernphysik, and
J\"{u}lich~Center~for~Hadron~Physics,~Forschungszentrum~J\"{u}lich,
D-52425~J\"{u}lich, Germany}
\affiliation{JARA~-~High~Performance~Computing, Forschungszentrum~J\"{u}lich, 
D-52425 J\"{u}lich,~Germany}

\author{Evgeny Epelbaum}\affiliation{Institut~f\"{u}r~Theoretische~Physik~II,~Ruhr-Universit\"{a}t~Bochum,
D-44870~Bochum,~Germany}

\author{Hermann~Krebs}
\affiliation{Institut~f\"{u}r~Theoretische~Physik~II,~Ruhr-Universit\"{a}t~Bochum,
D-44870~Bochum,~Germany}

\author{Gautam~Rupak}
\affiliation{Department~of~Physics~and~Astronomy, Mississippi~State~University, Mississippi State, MS~39762, USA}

\date{\today}

\begin{abstract}
Projection Monte Carlo calculations of lattice Chiral Effective Field Theory suffer from sign oscillations to a varying degree dependent on
the number of protons and neutrons. Hence, such studies have hitherto been concentrated on nuclei with equal numbers of protons and
neutrons, and especially on the alpha nuclei where the sign oscillations are smallest. Here, we introduce the ``symmetry-sign
extrapolation'' method, which allows us to use the approximate Wigner SU(4) symmetry of the nuclear interaction to systematically extend the 
Projection Monte Carlo calculations to nuclear systems where the sign problem is severe.
We benchmark this method by calculating the ground-state energies of the $^{12}$C,  
$^6$He and $^6$Be nuclei, and discuss its potential for studies of neutron-rich halo nuclei and asymmetric nuclear matter.
\end{abstract}

\pacs{21.10.Dr, 21.30.-x, 21.60.De}
\maketitle

\section{Introduction\label{sect:intro}}

Lattice Chiral Effective Field Theory~(EFT) is an {\it ab initio} framework~\cite{Lee:2004si,Borasoy:2006qn,Borasoy:2007vi,Epelbaum:2009zsa,Lee:2008fa}
which has recently been applied to studies of the structure of light and medium-mass nuclei~\cite{Epelbaum:2009pd,Epelbaum:2010xt,Lahde:2013uqa,Lahde:2013kma}, 
as well as to the physics of dilute neutron matter~\cite{Borasoy:2007vk,Epelbaum:2008vj}.
In particular, the structure of $^{12}$C and $^{16}$O has recently been elucidated using lattice 
Chiral EFT~\cite{Epelbaum:2012qn,Epelbaum:2013paa,Lahde:2014bna}. The role of the Hoyle state in $^{12}$C has also been investigated,
along with its anthropic implications for the viability of carbon-based life as we know it~\cite{Epelbaum:2012iu,Epelbaum:2013wla}. 
These successes notwithstanding, lattice Chiral EFT has so far mainly been
applied to ``alpha nuclei'', {\it i.e.} to nuclei with $A$ a multiple of 4 and with an equal number of protons and neutrons. This limitation is due to the
appearance of sign oscillations (the so-called sign problem) in the Projection Monte Carlo~(PMC) calculation at large Euclidean time. The sign problem also
necessitates the use of relatively large lattice spacings of $a \simeq 2$~fm, in order to moderate the repulsive short-range contributions in the leading-order~(LO) lattice 
Chiral EFT Hamiltonian. Even then, the low-lying spectra of most alpha nuclei can only be extracted after considerable extrapolation in Euclidean time. For instance, 
this has so far precluded studies of neutron-rich halo nuclei where sign oscillations are more severe.

In spite of the sign problem prevalent in lattice Chiral EFT, useful results have been made possible by the observation that nuclei can be approximately
described by a Hamiltonian which respects the Wigner SU(4) symmetry where spin and isospin degrees of freedom are interchangeable~\cite{Wigner:1936dx}.
Since Euclidean time projection with an SU(4) symmetric Hamiltonian is possible for most nuclei without a sign problem, one can obtain a trial wave function which is much closer to
the ground state of the full Hamiltonian. The shorter Euclidean projection time possible with the full Hamiltonian then becomes sufficient, and moreover the coupling constant
of the SU(4) symmetric Hamiltonian can be varied in order to generate a large set of independent trial states. Such self-consistent extrapolation in Euclidean time, also
referred to as ``triangulation'', allows for a much more precise determination of the properties of the nuclei under investigation than would otherwise be possible.
The usefulness of the Wigner SU(4) symmetry was also noted earlier in a different but related context. It was shown in Ref.~\cite{Kaplan:1995yg}, 
that the nuclear interactions are SU(4) symmetric in the limit of infinite colors, $N_C^{} \to \infty$. Further insight was obtained in Ref.~\cite{Mehen:1999qs}, where it was 
found that in the limit of large $S$-wave scattering lengths, the two-nucleon interactions exhibit Wigner SU(4) symmetry.

We shall now take the idea of the SU(4) symmetric auxiliary Hamiltonian one step further by considering a weighted sum of the full and SU(4) symmetric Hamiltonians.
In particular, the weight of each component in the Hamiltonian is controlled by a parameter $d_h^{}$ such that for $d_h^{} = 1$ we have the full Hamiltonian only, and
for $d_h^{} = 0$ we have the SU(4) symmetric Hamiltonian only. For values $0 \leq d_h^{} \leq 1$, we have a linear combination of the two Hamiltonians. The
properties of the physical system are then found by extrapolation to $d_h^{} \to 1$, combined with an extrapolation in Euclidean time. As will be shown here, this allows
us to circumvent the sign problem to a large extent, and opens up the possibility to study neutron-rich nuclei and systems where $N \neq Z$. It also allows us to access much
larger Euclidean projection times with the physical Hamiltonian, thereby greatly increasing the level of confidence in the Euclidean time extrapolation. We shall demonstrate
this ``symmetry-sign extrapolation'' (SSE) method by refining our earlier results for the ground state of $^{12}$C, and by computing the ground state 
energies of $^6$He and $^6$Be.

We begin in Section~\ref{sect:symm} by introducing the SSE method and discussing the origin of the sign problem in lattice Chiral EFT, and in 
Section~\ref{sect:12C} we present our method of analyzing the PMC data, along with our updated results for $^{12}$C. 
In Section~\ref{sect:6He}, we apply symmetry-sign extrapolation to the $^6$He nucleus and its mirror isobar $^6$Be. Finally, in Section~\ref{discussion} we 
discuss the prospects for extending our method to more neutron-rich nuclei and to nuclei where $N \neq Z$.

\section{The symmetry-sign extrapolation method\label{sect:symm}}

In order to introduce the SSE method, we first briefly recall the ingredients of lattice Chiral EFT at LO in the EFT expansion 
in $Q/\Lambda$. The LO contribution is treated non-perturbatively, while NLO and higher order terms are included as a perturbative correction. 
We use the same notation as in Ref.~\cite{Epelbaum:2009zsa}, and further details of the lattice action can be found there. At LO, the lattice Chiral EFT partition function is given by
\begin{eqnarray}
{\mathcal Z}_\mathrm{LO}^{} &=&
\int {\mathcal D}\pi_I^\prime  \,
\exp\left[ -S_{\pi\pi}^{}(\pi_I^\prime)  \right] \: \mathrm{Tr} \bigg\{ M^{}_\mathrm{LO}(\pi^\prime_I,N_t^{}-1)
\cdots \, M^{}_\mathrm{LO}(\pi^\prime_I,0) \bigg\}, 
\end{eqnarray}
where $\pi_I^\prime$ is the pion field and $S_{\pi\pi}^{}$ is the free pion lattice action.  As in Ref.~\cite{Epelbaum:2009zsa}, we are using an improved LO operator where the contact interactions 
depend on the momentum transfer through a smooth smearing function $f(\vec{q})$. The normal-ordered LO transfer matrix operator at Euclidean time step $n_t^{} = 0, \ldots, N_t^{}$ is
\begin{eqnarray}
&& M^{(n_t)}_\mathrm{LO}(\pi^\prime_I)
 = \: : \exp\left[-H_\mathrm{LO}^{}(\pi^\prime_I,n_t^{})\alpha_t^{}
 \label{trans_1a}\right] :,
\end{eqnarray}
with
\begin{eqnarray}
H^{(n_t)}_\mathrm{LO}(\pi^\prime_I) &=& H_\mathrm{free}^{} + \frac{1}{2} C \, \sum_{\vec q} f(\vec q) \, : \rho(\vec q)\rho(-\vec q) : 
+ \: \frac{1}{2}C_{S^{2}}^{} \sum_{\vec q,S} f(\vec q)\, : \rho_S^{}(\vec q)\rho_S^{}(-\vec q) : \nonumber \\
&& + \: \frac{1}{2}C_{I^{2}}^{} \sum_{\vec q,I} f(\vec q): \rho_I^{}(\vec q)\rho_I^{}(-\vec q) :
+ \: \frac{1}{2}C_{S^{2},I^2}^{} \sum_{\vec q,S,I} f(\vec q) \, : \rho_{S,I}^{}(\vec q)\rho_{S,I}^{}(-\vec q) : \nonumber \\
&& + \: \frac{g_A^{}}{2f_\pi^{}\sqrt{q_\pi^{}}} \sum_{\vec n,S,I} \Delta_S^{} \pi^\prime_I(\vec n,t) \, \rho_{S,I}^{}(\vec n),
\label{trans_1b}
\end{eqnarray}
where the spin vector $S$ and the isospin vector $I$ indices range from 1 to 3, the parameter $\alpha_t^{} \equiv a_t^{}/a$ is the ratio of temporal and spatial lattice spacings, 
$\Delta_S^{}$ is the lattice gradient along direction $S$, and $q_\pi^{} \equiv \alpha_t^{} (m_\pi^2 + 6)$. The operator $\rho$ is the total nucleon density $N^{\dagger}N$, $\rho_S^{}$ is 
the spin density $N^{\dagger}\sigma_S^{} N$, $\rho_I^{}$ is the isospin density $N^{\dagger}\tau_I^{} N$, and $\rho_{S,I}^{}$ is the isospin density $N^{\dagger}\sigma_S^{}\tau_I^{} N$.
The couplings $C$, $C_{S^2}^{}$, $C_{I^2}^{}$ and $C_{S^2,I^2}^{}$ satisfy
\begin{align}
C &=-3C_{S^2,I^2}=-\frac{3}{2}(C_{S^2}+C_{I^2}),
\end{align}
such that the smeared contact interactions only contribute to even-parity channels, where we have antisymmetry in spin 
and symmetry in isospin (or {\it vice versa}). 

The PMC calculations are performed with auxiliary fields coupled to each of the densities $\rho$, $\rho_S^{}$, $\rho_I^{}$, and $\rho_{S,I}^{}$. 
The LO auxiliary-field transfer matrix at Euclidean time step $n_{t}^{}$ is
\begin{align}
M^{(n_{t})}_{\rm{LO,aux}}(s,s_{S}^{},s_{I}^{},s_{S,I}^{},\pi_{I}^{\prime})  & = \: : \exp\left\{
- H_{\text{free}}\alpha_{t}^{} 
- \frac{g_{A}\alpha_{t}^{}}{2f_{\pi}\sqrt{q_{\pi}^{}}} \sum_{\vec{n},S,I} \Delta_{S}^{}\pi_{I}^{\prime}(\vec{n},n_{t}^{})\rho_{S,I}^{}(\vec n)\right. \nonumber \\
+\: \sqrt{-C\alpha_{t}^{}} & \sum_{\vec n} s(\vec n,n_{t}^{})\rho(\vec{n}) 
+\: i \sqrt{C_{S^2}^{}\alpha_{t}^{}} \sum_{\vec{n},S}s_{S}^{}(\vec{n},n_{t}^{})\rho_{S}^{}(\vec{n}) \nonumber \\
+\: i \sqrt{C_{I^2}^{}\alpha_{t}^{}} & \sum_{\vec{n},I} s_{I}^{}(\vec{n},n_{t}^{})\rho_{I}^{}(\vec{n}) \left.  
+\: i \sqrt{C_{S^2,I^2}^{}\alpha_{t}^{}}\sum_{\vec{n},S,I} s_{S,I}^{} (\vec{n},n_{t}^{})\rho_{S,I}^{}(\vec{n})\right\} : \, ,
\label{aux_trans}
\end{align}
where the physical values of the LO operator coefficients are such that all factors inside the square root symbols in Eq.~(\ref{aux_trans}) are positive.
We now also define an SU(4) symmetric transfer matrix,
\begin{eqnarray}
&& M_4^{(n_t)} = \: : \exp\left[- H_4^{}\alpha_t^{} \right] :, 
\label{trans_4a}
\end{eqnarray}
with
\begin{eqnarray}
&& H_4^{} = H_\mathrm{free}^{} +\frac{1}{2}C_{4}^{} \, \sum_{\vec q} f(\vec q) \, : \rho^{}_{}(\vec q)\rho^{}_{}(-\vec q): \,, 
\label{trans_4}
\end{eqnarray}
where $C_4^{} < 0 $ is the associated coupling constant. Again, we can rewrite the interaction using auxiliary fields,
giving
\begin{equation}
M^{(n_{t})}_{\rm{4,aux}}(s) = \, : \exp\left\{
-H_{\rm{free}}\alpha_{t}  +\sqrt{-C_4\alpha_{t}} \sum_{\vec{n}}s(\vec{n},n_{t})\rho(\vec{n})\right\} \, : \, .
\end{equation}

Let us now consider the projection amplitude we obtain for an $A$-body Slater determinant initial state
\begin{align}
\left| \Psi \right>=\left| \psi_1 \right>\wedge\left| \psi_2 \right>\wedge \cdots \wedge\left| \psi_A \right>,
\end{align}
with an auxiliary-field transfer matrix $M^{(n_{t})}_{\rm{aux}}$. The projection amplitude is given by $\det({\bm M})$, where ${\bm M}$ is the $A \times A$ matrix we obtain from the 
single nucleon amplitudes
\begin{align}
{\bm M}_{i,j}= \left< \psi_i \right| M^{(N_{t})}_{\rm{aux}} \cdots M^{(0)}_{\rm{aux}}\left| \psi_j \right>.
\end{align}
Let us define ${\cal U}_*[M]$ as the set of unitary matrices such that $U^\dagger M U = M^*$. It can be shown that $\det({\bm M})$ is positive semi-definite if there exists some 
antisymmetric matrix $U \in {\cal U}_*[M]$. The proof follows from the fact that the spectra of $M$ and $M^*$ must coincide, and the real spectrum of $M$ must be doubly degenerate
as a result of the antisymmetry of $U$~\cite{Lee:2004ze}. A straightforward generalization of this result is that the projection amplitude $\det({\bm M})$ is positive semi-definite 
if there exists a unitary operator $U \in {\cal U}_*[M^{(n_{t})}_{\rm{aux}} ]$ for all $n_t^{} = 0,\cdots, N_t^{}$ and if the action of $U$ on the single-particle states 
$\left| \psi_1 \right>$, $\cdots$, $\left| \psi_A \right>$ can be represented as an antisymmetric $A\times A$ matrix.

We note that ${\cal U}_*[M^{(n_{t})}_{4,\rm{aux}}]$ contains the spin and isospin matrices $\sigma_2^{}$ and $\tau_2^{}$. PMC calculations
with the SU(4)-symmetric theory are then free from sign oscillations whenever the initial single nucleon states are paired into spin singlets or isospin 
singlets~\cite{Chen:2004rq,Lee:2007eu}. 
We may then define the ``interpolating Hamiltonian'' $H$ as
\begin{equation}
H \equiv d_h^{} H_\mathrm{LO}^{} + (1-d_h^{}) H_4^{},
\label{Hd}
\end{equation}
which depends on $d_h^{}$ as well as the (unphysical) coupling constant $C_4^{}$. This can also be viewed as giving the interaction parameters a linear dependence on $d_h^{}$,
\begin{align}
C(d_h^{}) &\equiv d_h^{} C + (1-d_h^{}) C_{4}^{}, \nonumber \\
C_{S^2}^{}(d_h^{}) &\equiv d_h^{} C_{S^2}^{}, \;\; 
C_{I^2}^{}(d_h^{})\equiv d_h^{} C_{I^2}^{}, \nonumber \\ \;\; 
C_{S^2,I^2}^{}(d_h^{}) &\equiv d_h^{} C_{S^2,I^2}^{}, \;\; g_A^{}(d_h^{})\equiv d_h^{} g_A^{}.
\end{align}
By taking $d_h^{} < 1$, we can always
decrease the sign problem to a tolerable level, while simultaneously tuning $C_4^{}$ to a value favorable for an extrapolation $d_h^{} \to 1$. Most 
significantly, we can make use of the constraint that the physical result at $d_h^{} = 1$ should be independent of $C_4^{}$. 
The dependence of calculated matrix elements on $d_h^{}$ is smooth in the vicinity of $d_h^{} = 1$. 
In what follows, we shall explore the properties of $H$ for various nuclei of physical interest, and determine to what extent it can
be used to circumvent the sign problem, which at $d_h^{} = 1$ becomes exponentially severe in the limit of large Euclidean time.  

We note that an extrapolation technique similar to SSE has been used in Shell Model Monte Carlo calculations for over two decades~\cite{Alhassid:1993yd,Koonin:1996xj}. 
In that case, the extrapolation is performed by decomposing the Hamiltonian into ``good sign'' and ``bad sign'' parts, $H_G^{}$ and $H_B^{}$, respectively. The calculations are then performed 
by multiplying the coefficients of $H_B^{}$ by a parameter $g$ and extrapolating from $g < 0$, where the simulations are free from sign oscillations, to the physical point $g = 1$.
For SSE, the analysis in terms of  ``good'' and ``bad'' signs is not the entire story. Most of the interactions can be divided into two groups which are ``sign free'' by themselves, such that a large 
portion of the sign oscillations is due to interference between the different underlying symmetries of the two groups of interactions. Since this effect is quadratic in the interfering interaction 
coefficients, the growth of the sign problem is more gradual. We therefore expect to be able to extrapolate from values not so far away from the physical point $d_h^{} = 1$.

The LO auxiliary-field transfer matrix $M^{(n_{t})}_{\rm{LO,aux}}$ contains pion interactions with the matrix structure $\sigma_S^{} \tau_I^{}$ acting on single nucleon states, and 
smeared contact interactions with matrix structures $\openone$, $i\sigma_S^{}$, $i\tau_I^{}$ and $i\sigma_S^{} \tau_I^{}$. 
Since
\begin{align}
\sigma_2^{} &\in {\cal U}_*[\sigma_S^{}\tau_2^{}],\, {\cal U}_*[i\sigma_S^{}], \, {\cal U}_*[i\tau_2^{}], \, {\cal U}_*[i\sigma_S^{}\tau_1^{}], \,
{\cal U}_*[i\sigma_S^{}\tau_3^{}], \label{sigma_2} \\
\sigma_2^{}\tau_3^{} &\in {\cal U}_*[\sigma_S^{}\tau_1^{}], \, {\cal U}_*[i\sigma_S^{}], \, {\cal U}_*[i\tau_1^{}], \, {\cal U}_*[i\sigma_S^{}\tau_2^{}],
\label{sigma_2tau_3}
\end{align}
we note that if we have an initial state with an even number of neutrons paired into spin-singlets and an even (but in general different) number of protons paired into spin-singlets, 
then there will exist an antisymmetric representation for both $\sigma_2^{}$ and $\sigma_2^{}\tau_3^{}$ on the single nucleon states.  
The only interaction matrix structures in $M^{(n_{t})}_{\rm{LO,aux}}$ not included in both Eqs.~(\ref{sigma_2}) and~(\ref{sigma_2tau_3}) are $i\tau_3^{}$
and $\sigma_S^{}\tau_3^{}$. Hence, the sign oscillations in $\det({\bm M})$ will be produced by $i\tau_3^{}$ and $\sigma_S^{}\tau_3^{}$, and the interference between 
the two sets of interactions in Eqs.~(\ref{sigma_2}) and~(\ref{sigma_2tau_3}). 
For initial states where the neutrons and protons cannot be paired into spin-singlets, there will be additional sign oscillations due to these unpaired nucleons. 
However, the number of such unpaired nucleons can often be kept to a minimum.

\section{Results for Carbon-12\label{sect:12C}}

We shall first discuss our results for the $^{12}$C nucleus, as this provides us with a convenient test case for the SSE method.
We extend the PMC calculation to larger values of the Euclidean projection time than would otherwise be possible, and verify how well these new results
agree with earlier calculations. Thus, we shall work at finite Euclidean time, extrapolate such data to $d_h^{} \to 1$, after which the extrapolation
$L_t^{} \to \infty$ is performed. For the case of the $^6$He nucleus, we shall consider the opposite order of limits and discuss in which cases either method
is preferable.

\begin{figure}[hb!]
\includegraphics[width=.45\linewidth]{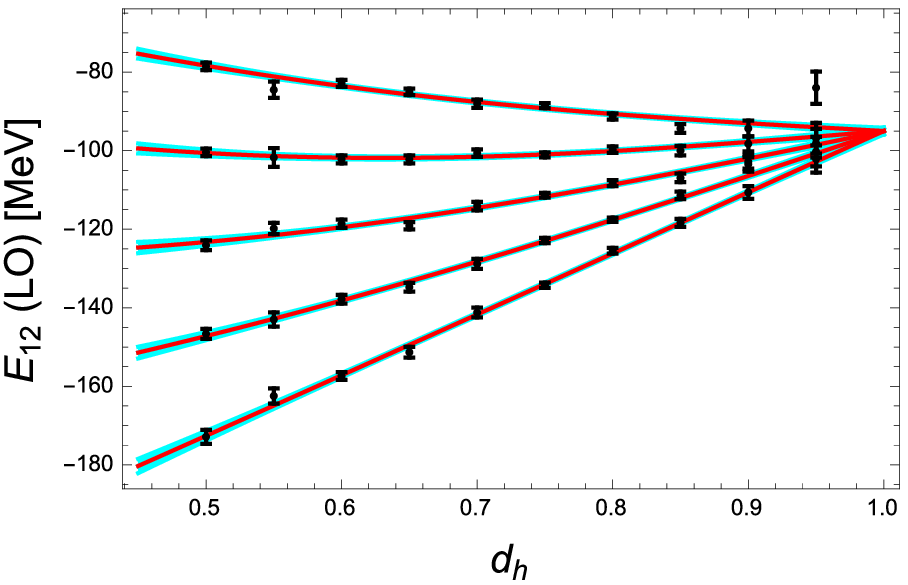}\quad\includegraphics[width=.45\linewidth]{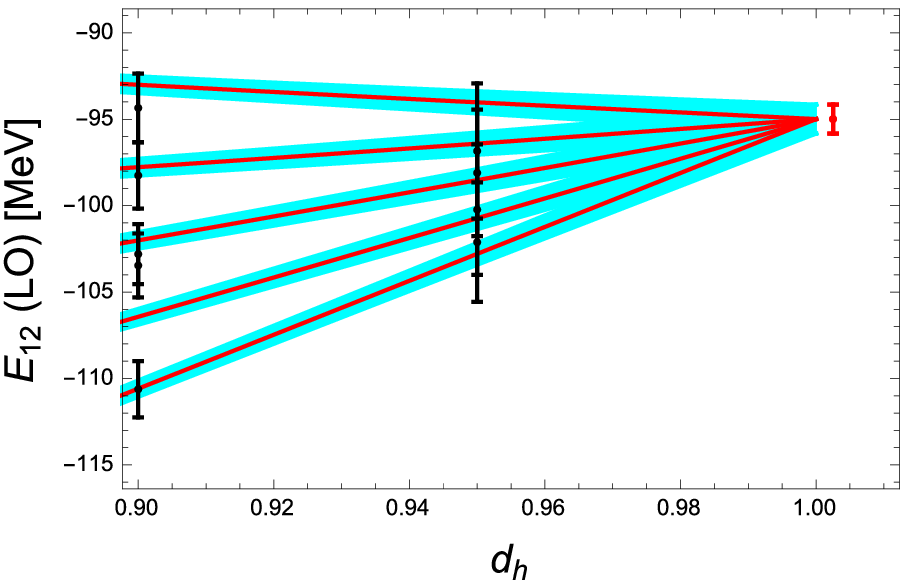}\\
\includegraphics[width=.45\linewidth]{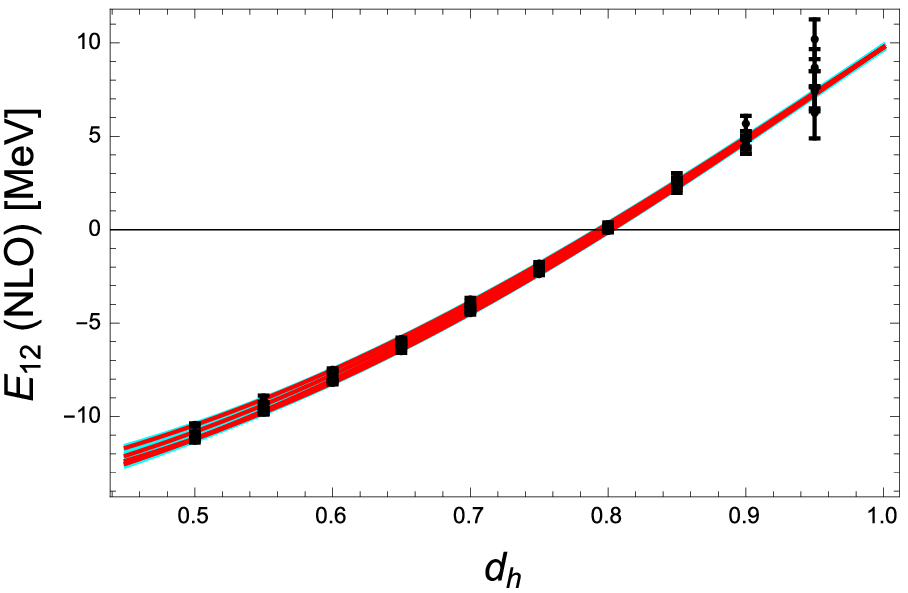}\quad\includegraphics[width=.45\linewidth]{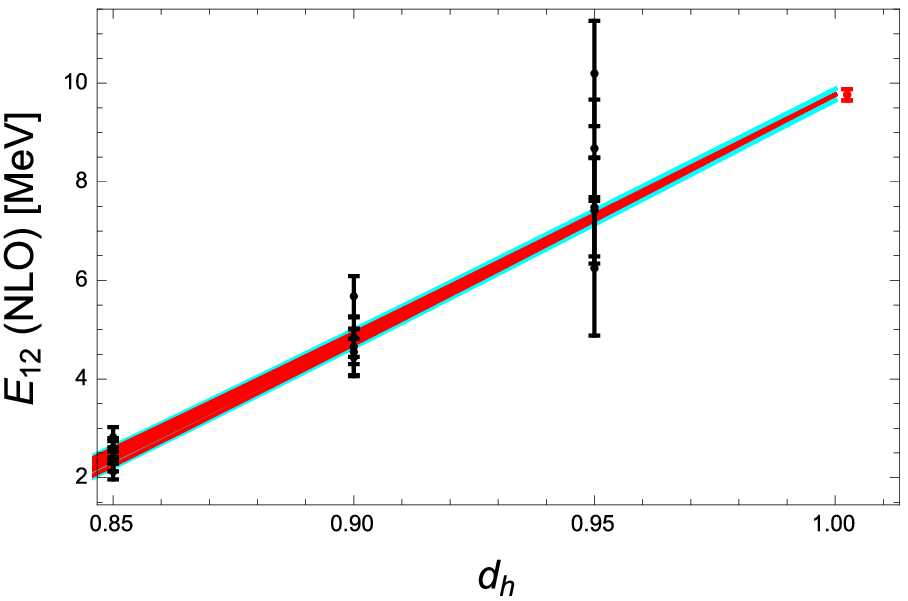}
\caption{\label{fig:12C Lt12 LO NLO} PMC data at $N_t^{} = 12.5$ for $^{12}$C, with the upper figures showing the LO energy and the lower figures
the shift due to two-nucleon forces at NLO. The left column shows the entire range in $d_h^{}$, while the right column shows a close-up around the physical point
$d_h^{} = 1$. The red lines are a simultaneous fit to all the PMC data using Eq.~(\ref{eqn:fit function}), and each line corresponds to the Hamiltonian~(\ref{Hd}) with a different
choice of the coupling constant $C_4^{}$. The result of the extrapolation $d_h^{} \to 1$ is given by the red points in the right column of plots, along with the statistical
uncertainty of the extrapolation. The error bars of the individual data points at $d_h^{} < 1$ represent the Monte Carlo uncertainties. The cyan error bands correspond to
the 67\% confidence levels of the extrapolations.} 
\end{figure}

Our strategy for data analysis is to perform a global fit to all PMC data obtained for different values of $d_h^{}$ and the coupling constant $C_4^{}$ of the
SU(4) symmetric Hamiltonian. The {\it ansatz} is
\begin{equation}
\label{eqn:fit function}
X(d_h^{},C_4^{},n) \equiv
X_0^{}+X_0^{\text{SU(4)}}(1-d_h^{})+\sum_{j=1}^n X_j^{\text{SU(4)}} \sin(j\pi d_h^{}),
\end{equation}
where the parameters $X_0^{}$ and $X_i^{\text{SU(4)}}$ are determined by a weighted least-squares fit. The superscript ``SU(4)'' implies that such 
parameters depend on $C_4^{}$. On the other hand, $X_0^{}$ represents the extrapolated value of the observable at
$d_h^{} = 1$, is therefore by definition independent of $C_4^{}$, and provides an important constraint for the least-squares fit. The determination of 
$X_0^{}$ for different observables provides the physical information of the analysis, and is obtained by a simultaneous fit to multiple instances of
the (unphysical) coupling constant $C_4^{}$. In general, the results for $d_h^{} \neq 1$ depend on the choice of $C_4^{}$.

The ``order parameter'' $n$ in Eq.~(\ref{eqn:fit function}) is adjusted to produce a $\chi^2$ per degree of freedom $\simeq 1$. For most observables, we find 
that $n=1$ is sufficient, with some notable exceptions that we shall describe later. For $n \ge 4$, the $\chi^2$ per degree of freedom quickly saturates below $1$ and our fits 
become over-constrained.  When different values of $n$ produce similar $\chi^2$ per degree of freedom, we choose the lowest order for which 
we perform the extrapolation $d_h^{} \to 1$. Our fitting function~(\ref{eqn:fit function}) is suggested by the expectation that the dependence of observables 
on $d_h^{}$ should, according to perturbation theory, be linear around $d_h^{} \simeq 0$ and $d_h^{} \simeq 1$. The linearity around $d_h^{} = 1$ is
incorporated by the sine function in Eq.~(\ref{eqn:fit function}).

\begin{figure}
\includegraphics[width=.45\linewidth]{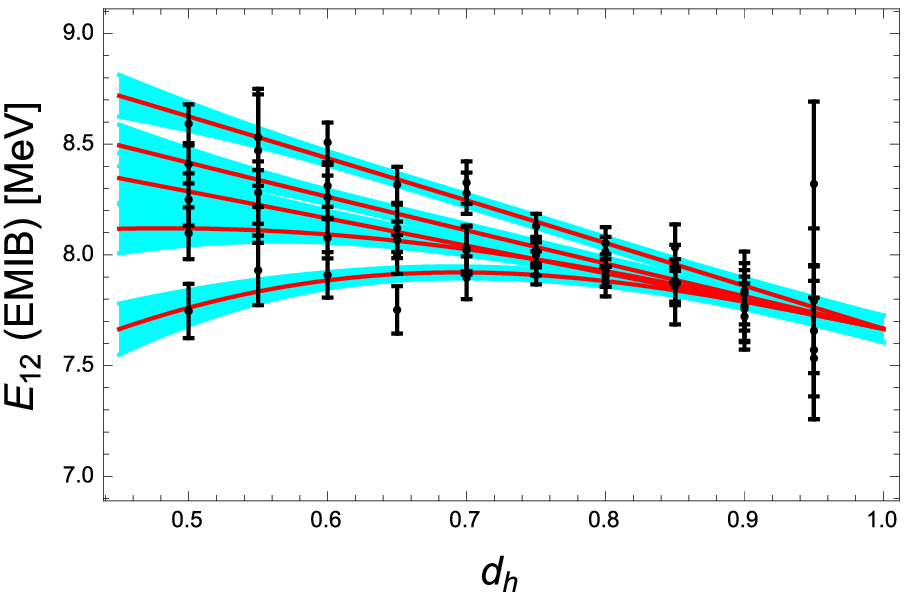}\quad\includegraphics[width=.45\linewidth]{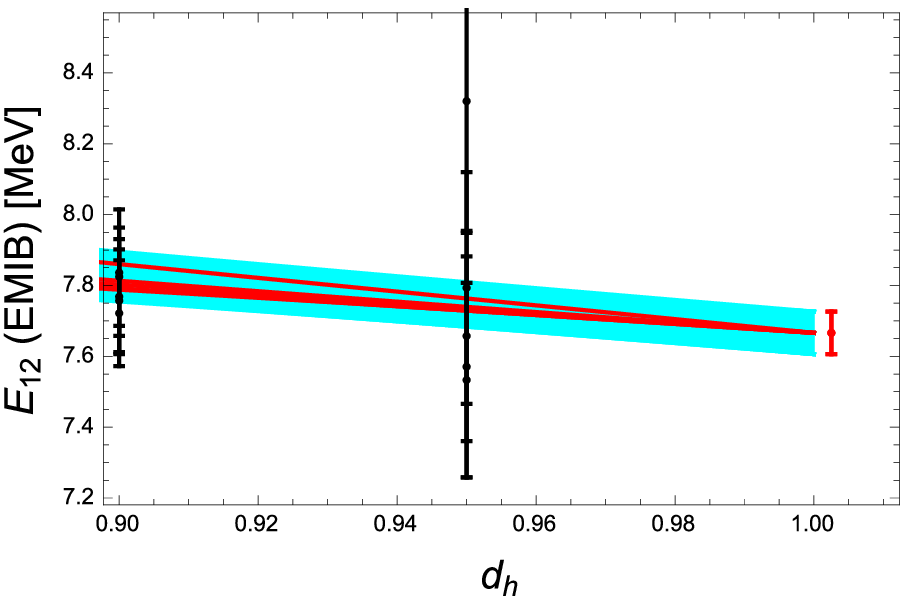}\\
\includegraphics[width=.45\linewidth]{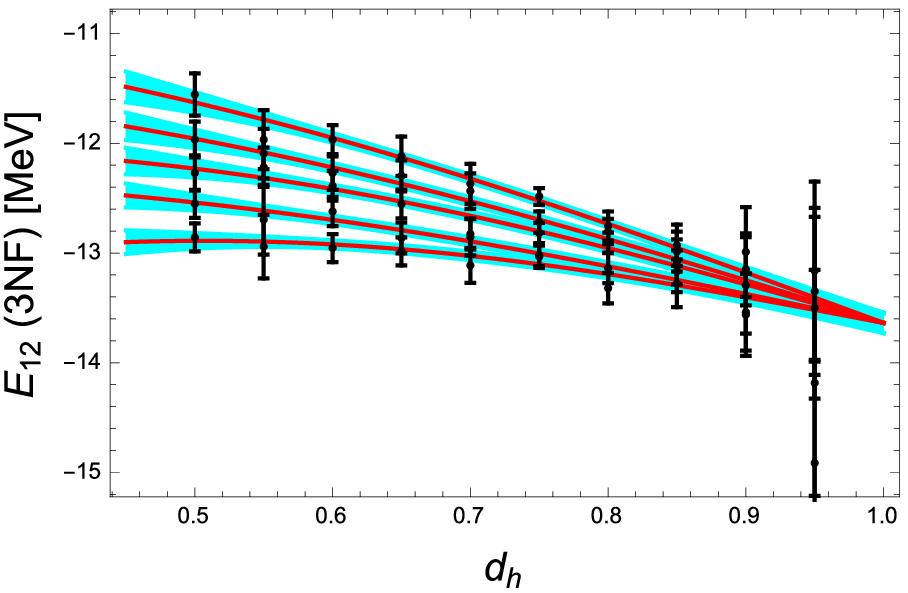}\quad\includegraphics[width=.45\linewidth]{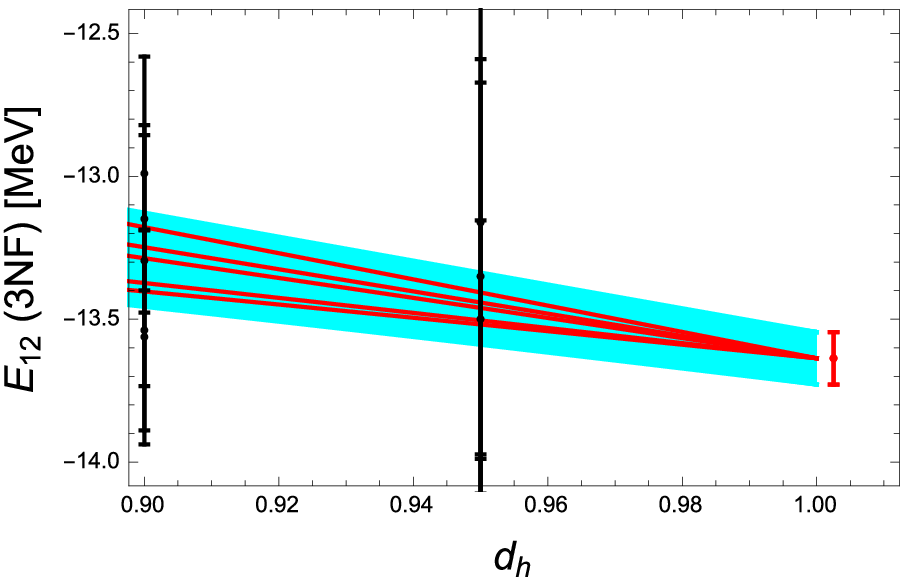}
\caption{\label{fig:12C Lt12 EMIB 3NF} PMC data at $N_t^{} = 12.5$ for $^{12}$C, with the upper figures showing the total contribution
from electromagnetic and strong isospin breaking (EMIB) operators, and the lower figures showing that of the three-nucleon force (3NF) operators
at NNLO. Notation and conventions are identical to Fig.~\ref{fig:12C Lt12 LO NLO}.}
\end{figure}

\begin{figure}
\includegraphics[width=.45\linewidth]{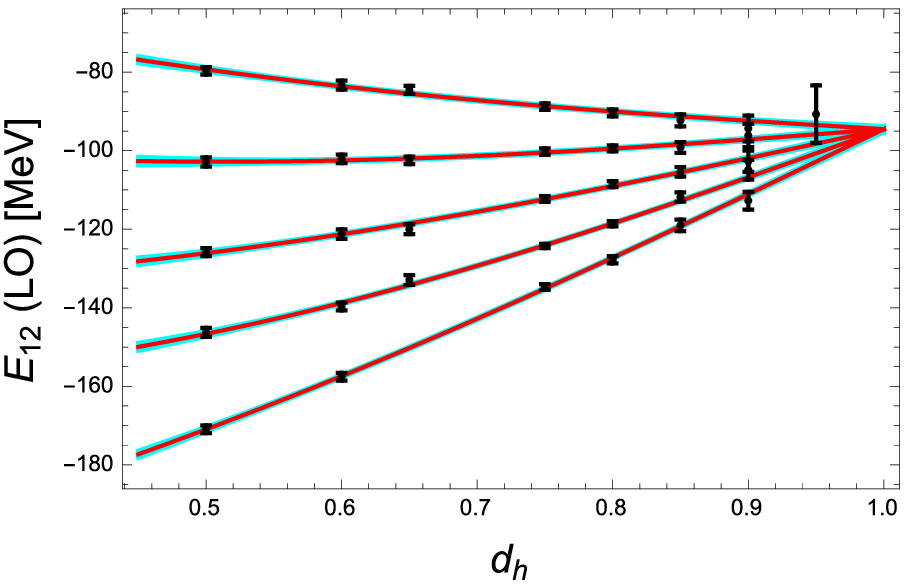}\quad\includegraphics[width=.45\linewidth]{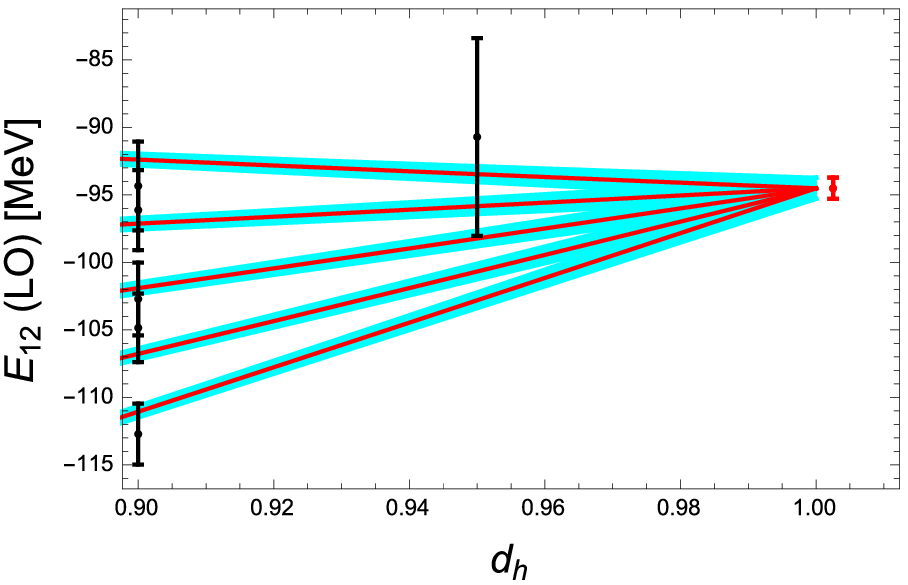}\\
\includegraphics[width=.45\linewidth]{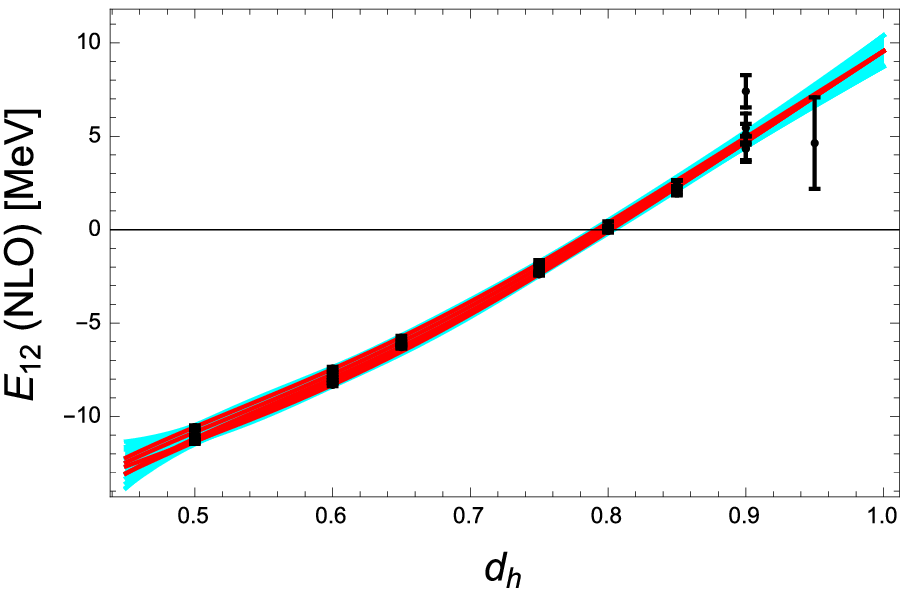}\quad\includegraphics[width=.45\linewidth]{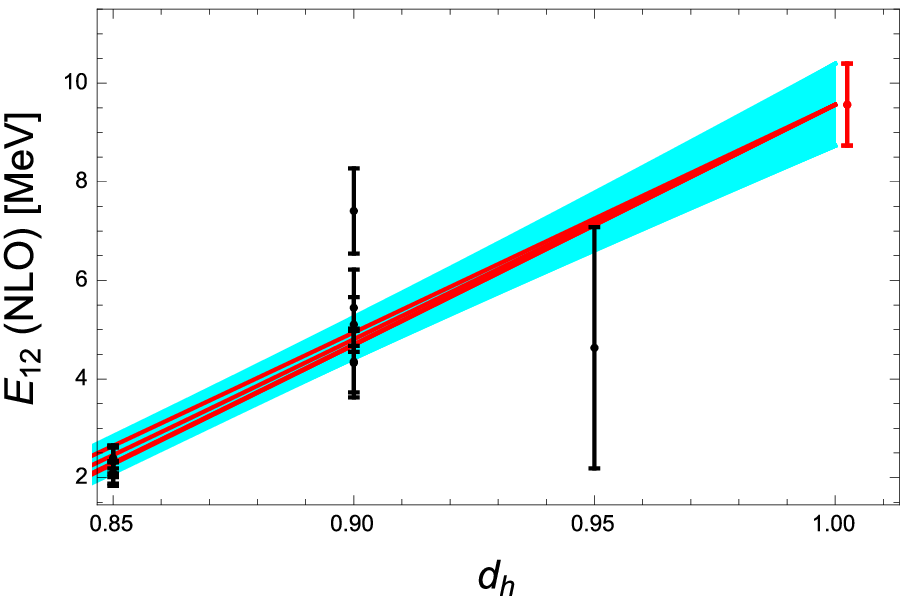}
\caption{\label{fig:12C Lt14 LO NLO} PMC data (LO and NLO contributions) at $N_t^{} = 14.5$ for $^{12}$C. For a full description of the
data and analysis, see the caption of Fig.~\ref{fig:12C Lt12 LO NLO}.} 
\end{figure}

\begin{figure}
\includegraphics[width=.45\linewidth]{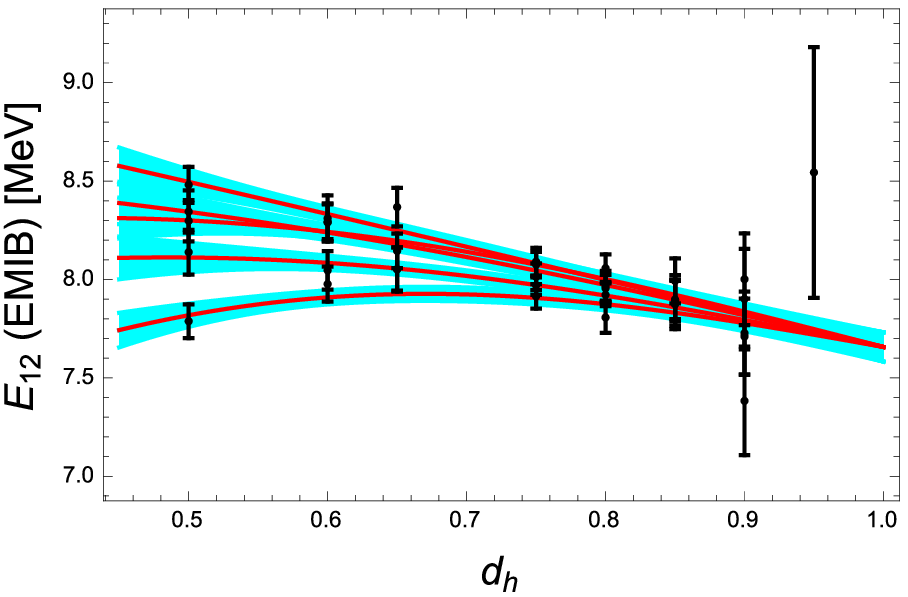}\quad\includegraphics[width=.45\linewidth]{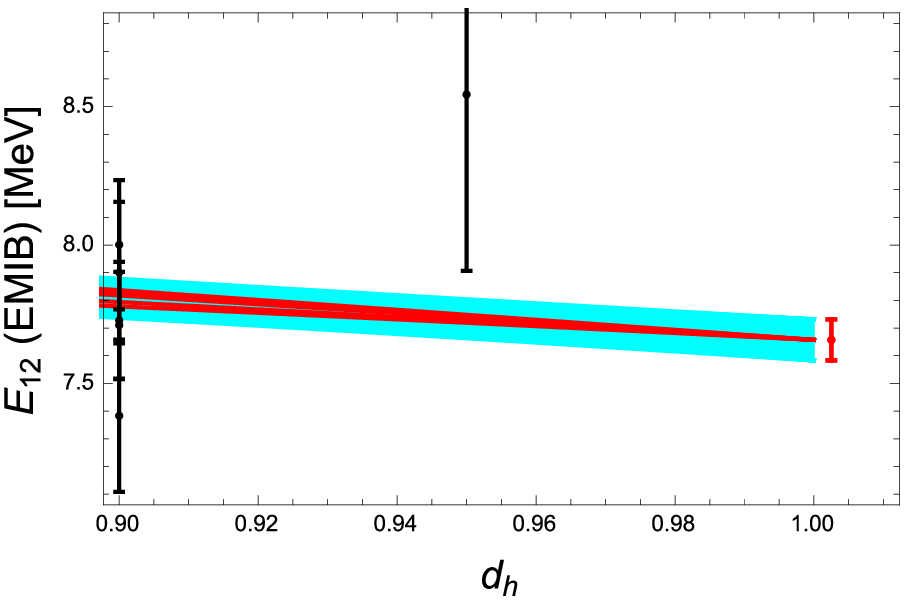}\\
\includegraphics[width=.45\linewidth]{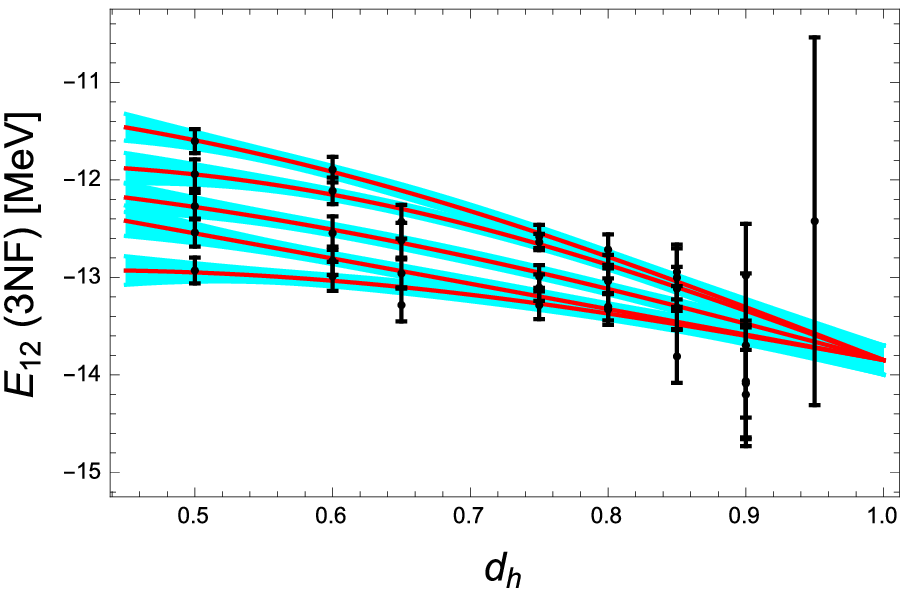}\quad\includegraphics[width=.45\linewidth]{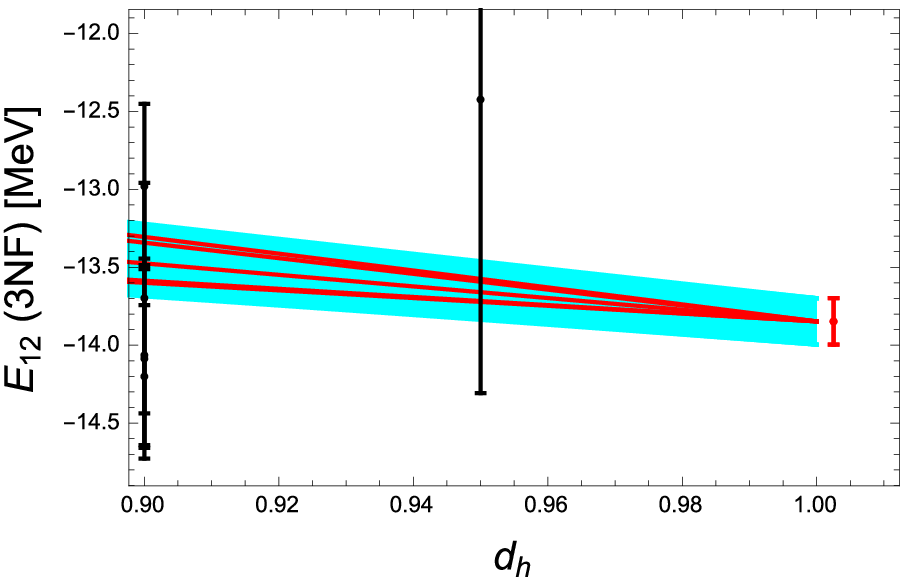}
\caption{\label{fig:12C Lt14 EMIB 3NF} PMC data (EMIB and 3NF contributions) at $N_t^{} = 14.5$ for $^{12}$C. For a full description of the
data and analysis, see the caption of Fig.~\ref{fig:12C Lt12 EMIB 3NF}.}
\end{figure}

In Figs.~\ref{fig:12C Lt12 LO NLO}-\ref{fig:12C Lt14 EMIB 3NF}, we show PMC data and extrapolations in $d_h^{}$ for the leading order~(LO) contribution, the next-to-leading order 
two-nucleon force~(NLO), the electromagnetic and strong isospin breaking~(EMIB) and the next-to-next-to-leading order three-nucleon force~(3NF), for the $^{12}$C ground state energy at
$N_t^{} = 12.5$ and $N_t^{} = 14.5$. The PMC data points in these figures are comprised of runs with different choices of $C_4^{}$ in the underlying SU(4) symmetric Hamiltonian. These
are $C_4^{} = -3.2 \times 10^{-5}$, $-3.4 \times 10^{-5}$, $-3.8 \times 10^{-5}$, $-4.2 \times 10^{-5}$ and $-4.8 \times 10^{-5}$ (in units of~MeV$^{-2}$), which can be distinguished as 
separate bands in the figures. The uppermost band corresponds to $C_4^{} = -3.2 \times 10^{-5}$~MeV$^{-2}$, and the lowest band to $C_4^{} = -4.8 \times 10^{-5}$~MeV$^{-2}$. 
With the exception of the NLO contribution, all fits were performed with
$n = 1$. For the NLO contribution, a higher order fit of $n = 3$ was used, in order to avoid introducing a bias around $d_h^{} \simeq 1$ due to the increasingly accurate data at small values of $d_h^{}$. 
The higher order fit function has sufficient freedom to fully account for the data at small $d_h^{}$.

From our PMC results, we find that even for large Euclidean projection times $N_t^{}$, the uncertainties of our calculations are well under control for $d_h^{} < 0.8$, which
shows that the sign problem is under control for such values of $d_h^{}$. However, as $d_h^{} \to 1$ the uncertainties clearly grow aggressively, such that the sign problem reaches full
strength at $d_h^{} = 1$. As previous work has demonstrated, PMC calculations for $^{12}$C at $d_h^{} = 1$ become very impractical for $N_t^{} > 10$. How far in $N_t^{}$ the
SSE analysis can be carried out depends on how robust the extrapolation in $d_h^{}$ can be made. We shall find that our extrapolated 
results remain robust as long as PMC calculations can be carried out for $d_h^{} > 0.75$, and that this range could possible be extended to smaller $d_h^{}$ by choosing $C_4^{}$ such that
the extent of the linear region around $d_h^{} = 1$ is maximized.

\begin{figure}
\includegraphics[width=\textwidth]{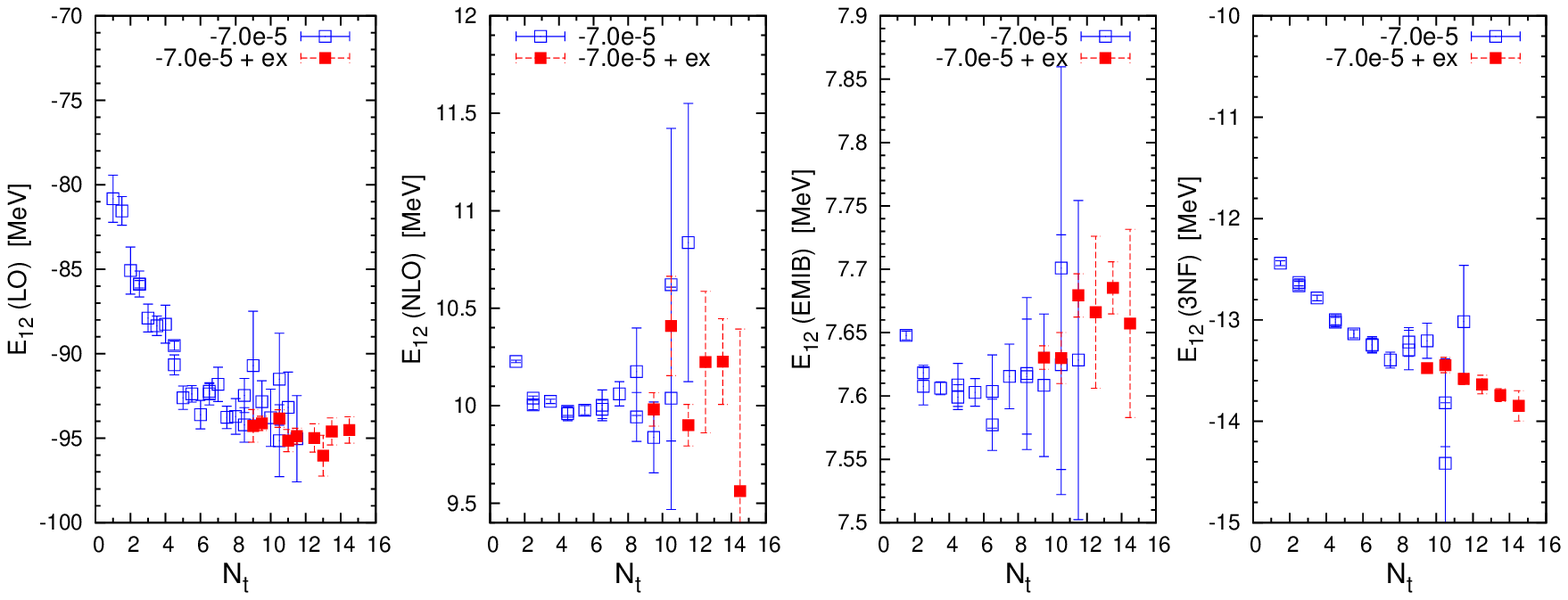}
\caption{\label{fig:12C_v11} Comparison of the new PMC data for $^{12}$C from the SSE analysis (red filled squares) and previous calculations~\cite{Lahde:2013uqa} 
for $d_h^{} = 1$ (blue open squares). The notation for the various contributions to the ground state energy $E_{12}^{}$ coincides with that of Table~\ref{tab:C12 results}. The results correspond
to a trial state with an SU(4) coupling of $-7.0 \times 10^{-5}$~MeV$^{-2}$, not to be confused with the SU(4) coupling $C_4^{}$ for the SSE analysis. It should be noted that the exponential
deterioration of the Monte Carlo error has been circumvented. Also, these data should not be interpreted in terms of a ``plateau'' as a function of $N_t^{}$. An analysis of the dependence on
$N_t^{}$ is given in Fig.~\ref{fig:12C main}, and a concise description of the Euclidean time extrapolation method can be found in Ref.~\cite{Lahde:2014sla}.}
\end{figure}
 
\begin{figure}
\includegraphics[width=\textwidth]{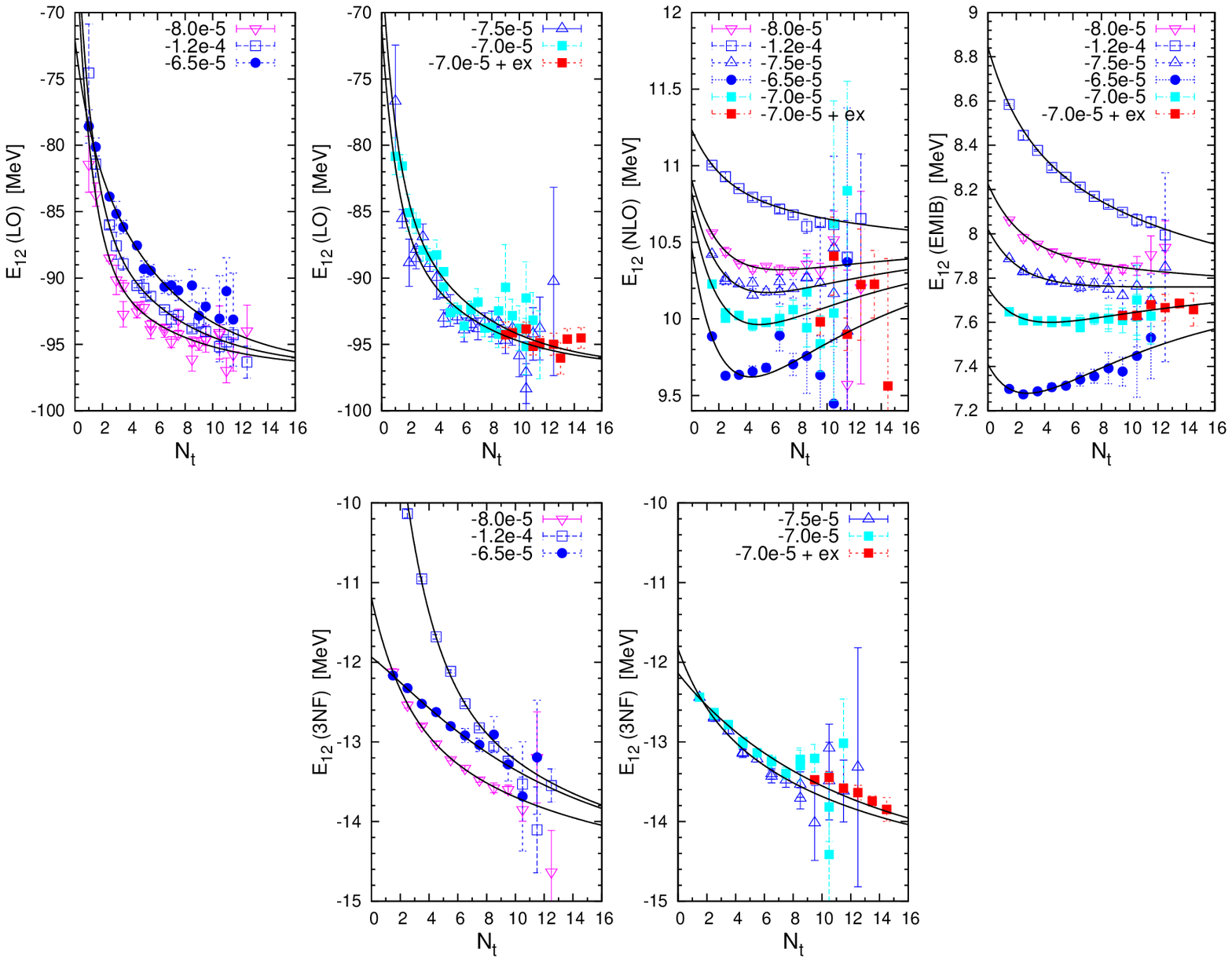}
\caption{\label{fig:12C main} Updated extrapolation in Euclidean projection time for the $^{12}$C ground state. The new results from this work (red filled squares) have been combined with
data compiled at $d_h^{} = 1$ from Ref.~\cite{Lahde:2013uqa}. The different data sets correspond to trial states with different SU(4) couplings (in units of~MeV$^{-2}$), 
not to be confused with the SU(4) coupling 
$C_4^{}$ for the SSE analysis. The results of the old and new analyses are given in Table~\ref{tab:C12 results}, and the extrapolation in Euclidean time is
discussed in detail in Ref.~\cite{Lahde:2014sla}.}
\end{figure}

For $^{12}$C, we have performed PMC calculations using SSE for Euclidean projection times between $N_t^{} = 9.0$ and~$14.5$ for one of the trial states
used in Ref.~\cite{Lahde:2013uqa}. It should be noted that the results of Ref.~\cite{Lahde:2013uqa} correspond to $d_h^{} = 1$, and could thus only be extended to $N_t^{} \simeq 10$ before the
sign problem became prohibitive. In Fig.~\ref{fig:12C_v11}, we extend the dataset of Ref.~\cite{Lahde:2013uqa} with the new SSE data, and in Fig.~\ref{fig:12C main} we 
combine all of our new results for $^{12}$C with those originally shown in Ref.~\cite{Lahde:2013uqa}. We have also repeated the extrapolation in Euclidean time with this updated data, with
a comparison and consistency check given in Table~\ref{tab:C12 results}. We find that our updated extrapolation is in agreement with the previous results of Ref.~\cite{Lahde:2013uqa}, and in most
cases with a slightly reduced uncertainty. While the reduction in uncertainty is small, this is due to the fact that our new dataset is quite limited in comparison to that used in Ref.~\cite{Lahde:2013uqa}.
Moreover, the main objective of our study of $^{12}$C is to establish that the SSE data are consistent with those obtained at $d_h^{} = 1$, when such calculations are not
prohibited by the sign problem.

\begin{table}
\caption{\label{tab:C12 results} Contributions to the ground state energy of $^{12}$C after extrapolation to infinite Euclidean projection time. The contributions from the improved leading order 
amplitude~(LO), the two-nucleon force at next-to-leading order~(NLO), the electromagnetic and strong isospin breaking~(EMIB) and the three-nucleon force at next-to-next-to-leading order~(3NF)
are shown separately. The left column shows the results using the PMC data for $d_h^{} = 1$ from Ref.~\cite{Lahde:2013uqa}, while the right column shows the results when the
SSE data from this work are included.}
\vspace{.2cm}
\begin{tabular}{l|rr}
& \multicolumn{1}{c}{Ref.~\cite{Lahde:2013uqa}} & \multicolumn{1}{c}{+ SSE} \\
\hline\hline
LO & $-96.92(16)$ & $-96.85(14)$ \\
NLO & $10.48(3)$ & $10.47(3)$ \\
EMIB & $7.76(1)$ & $7.76(1)$ \\
3NF & $-14.80(6)$ & $-14.56(4)$
\end{tabular}
\end{table}

\begin{figure}[ht!]
\includegraphics[width=.45\linewidth]{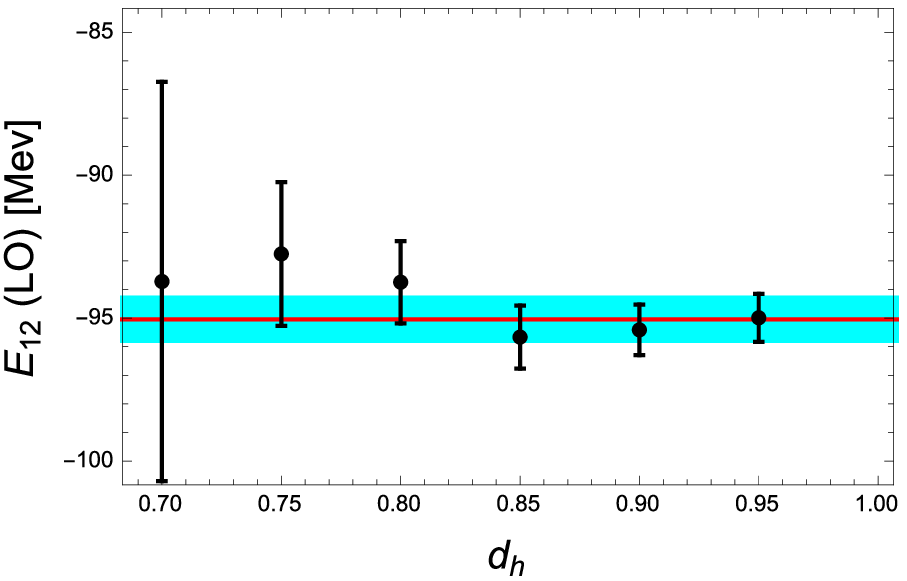} \quad \includegraphics[width=.45\linewidth]{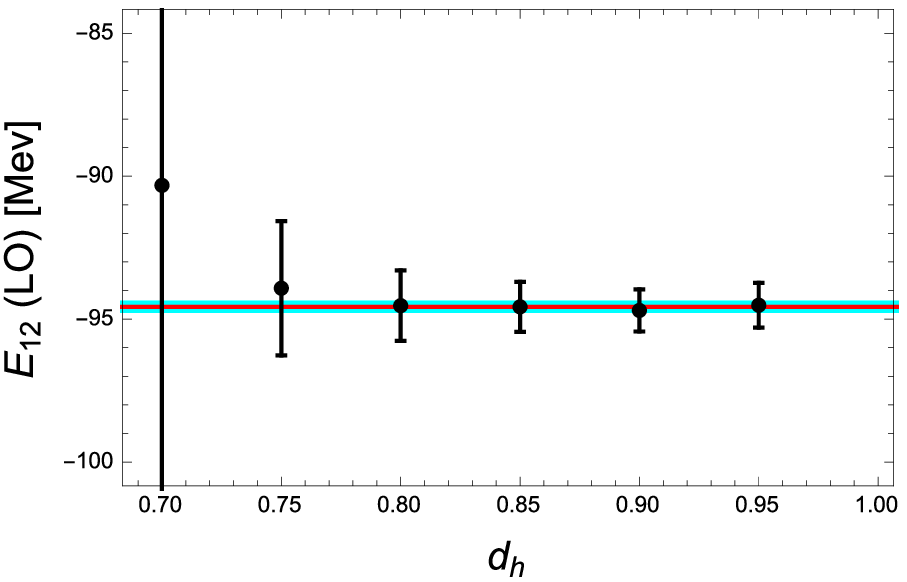}
\caption{\label{fig:C12 LO stability} Stability of the $^{12}$C ground state energy at LO at $N_t^{} = 12.5$ (left panel) and $N_t^{} = 14.5$ (right panel) under extrapolation $d_h^{} \to 1$. A data point
placed at a given value of $d_h^{}$ denotes that only data with equal and smaller values of $d_h^{}$ are included in the extrapolation. Thus, a
data point at $d_h^{} = 0.95$ only includes data with $d_h^{} \le 0.95$. The horizontal bands with 90\% confidence levels and are provided for visual guidance only.}
\end{figure}

It is worthwhile to investigate the stability of our extrapolated results as successive data points near $d_h^{} = 1$ are removed. In Fig.~\ref{fig:C12 LO stability}, we show the results of such a study 
for the ground state energy of $^{12}$C at LO. Each data point in Fig.~\ref{fig:C12 LO stability} corresponds to an extrapolation $d_h^{} \to 1$, such that the placement of the data point on the
horizontal axis indicates the ``break point'' at which data is excluded from the analysis. For example, for an extrapolated value placed on the horizontal axis at $d_h^{} = 0.8$, all PMC data
with $d_h^{} > 0.8$ have been excluded. For guidance, we also show a constant fit of the extrapolated points with 90\% confidence bands. The left panel of Fig.~\ref{fig:C12 LO stability}
shows data at $N_t^{} = 12.5$, while the right panel corresponds to $N_t^{} = 14.5$. In both cases, the extrapolated points located on the horizontal axis at $d_h^{} = 1$ are the ones shown 
in Fig~\ref{fig:12C main}.

A determination of the smallest value of $d_h^{}$ from which a reliable extrapolation $d_h^{} \to 1$ can be expected is highly significant, as systems with more nucleons and unequal numbers of
protons and neutrons will suffer from an increasingly severe sign problem. In order to compensate, this forces PMC calculations to be performed at successively smaller $d_h^{}$ (we note that
for $^{12}$C, the sign problem is already apparent in the larger uncertainties at $d_h^{} = 0.95$). As more complex nuclear systems are studied, the extent in which the extrapolated results can 
be ascertained from lower values of $d_h^{}$ will become more of an issue. Studying this behavior with our $^{12}$C data provides an initial rough idea on the robustness of the 
SSE method. As can be seen from Fig.~\ref{fig:C12 LO stability}, our extrapolated result is in good agreement for values of $d_h^{}$ as low as $\simeq 0.8$. 
Extrapolations using data below this value only become increasingly unreliable. This suggests the range in applicability of the dial parameter is limited. We find a similar conclusion within the 
PMC calculations of $^6$He, which we shall turn to next.

\section{Results for Helium-6 and Beryllium-6 \label{sect:6He}}

\begin{figure}
\includegraphics[width=.45\linewidth]{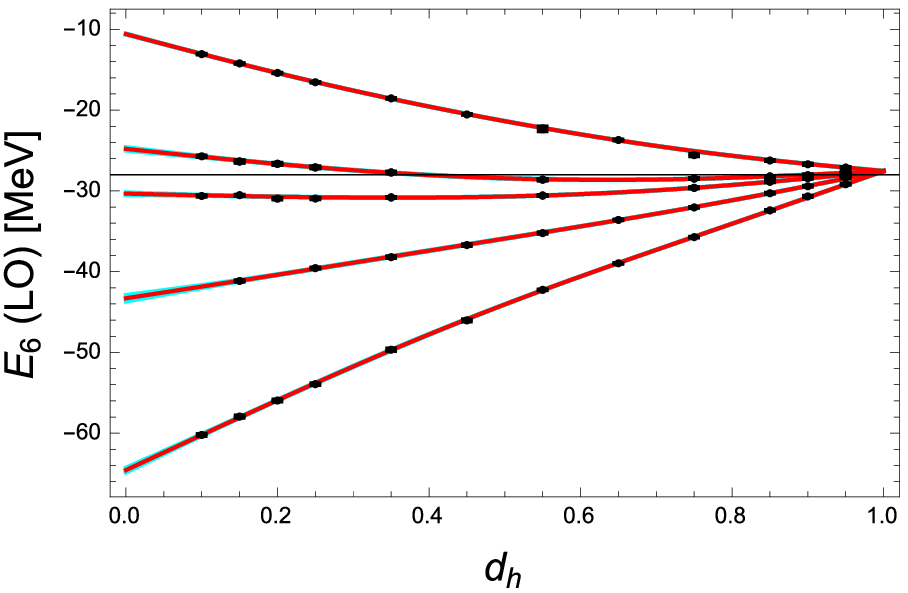}\quad\includegraphics[width=.45\linewidth]{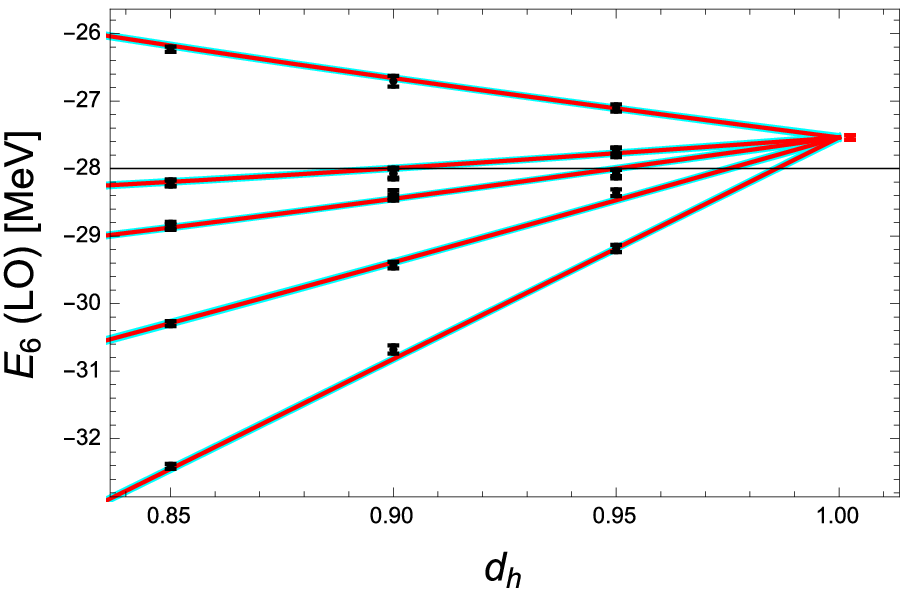} 
\\ \vspace{.3cm}
\includegraphics[width=.45\linewidth]{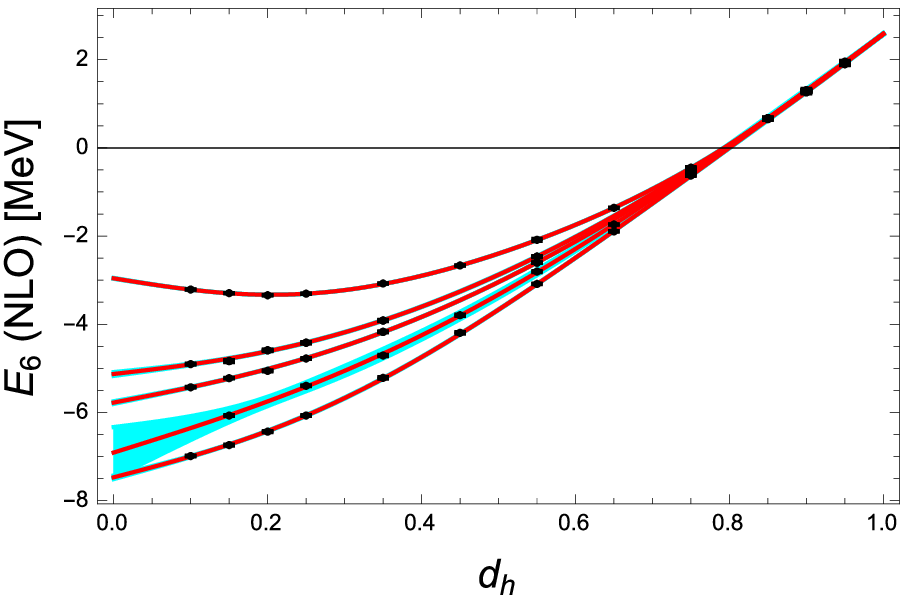}\quad\includegraphics[width=.45\linewidth]{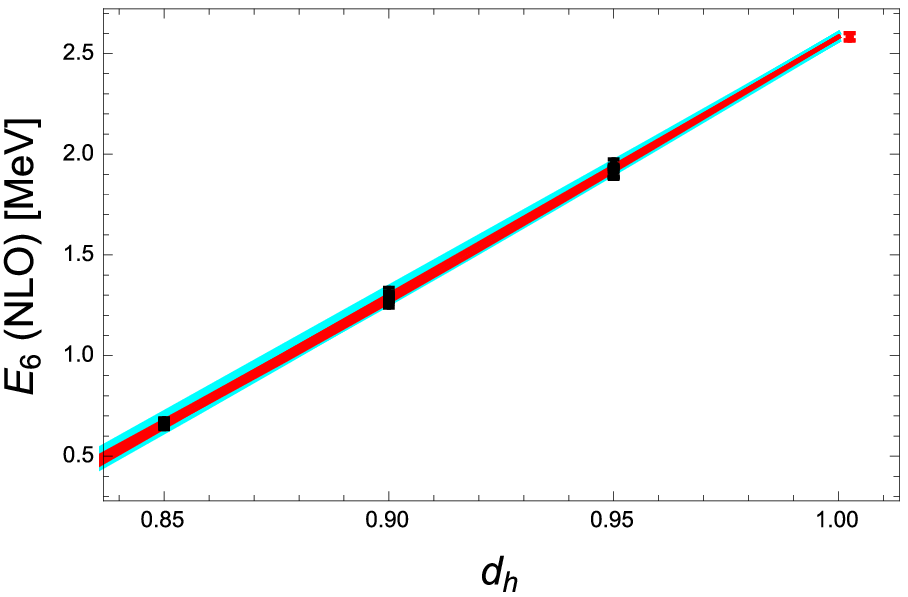}
\caption{\label{fig:He6 LO NLO} PMC results for the ground state energy of $^6$He at LO (upper panels) and the contribution from two-nucleon interactions at NLO (lower panels).
The left panels show the full range between $d_h^{} = 0$ and $d_h^{} = 1$, while the right panels show a close-up near $d_h^{} = 1$. Each data point has been individually extrapolated
to infinite Euclidean projection time before the SSE extrapolation $d_h^{} \to 1$. The results of the SSE extrapolation are shown by red data points in the right panels.}
\end{figure}

\begin{figure}
\includegraphics[width=.45\linewidth]{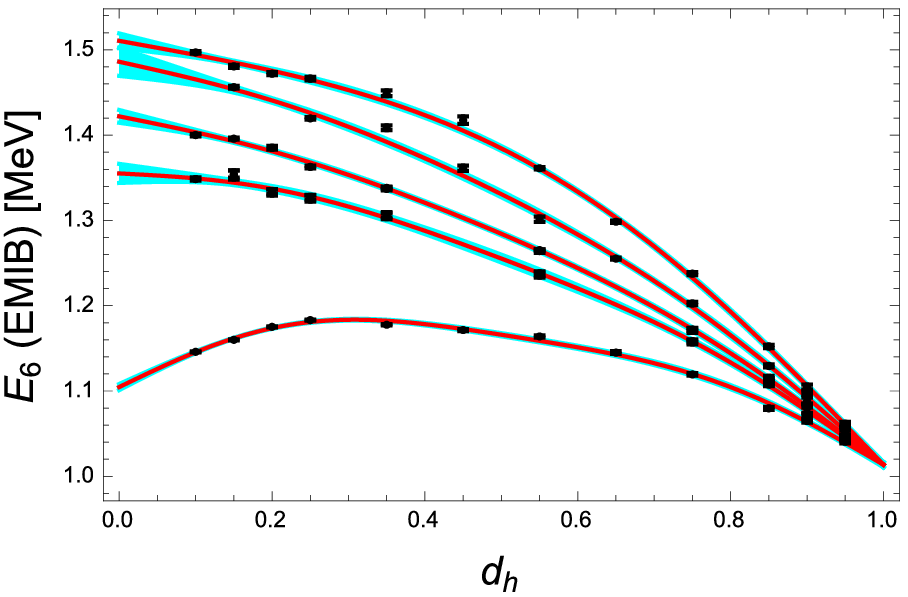}\quad\includegraphics[width=.45\linewidth]{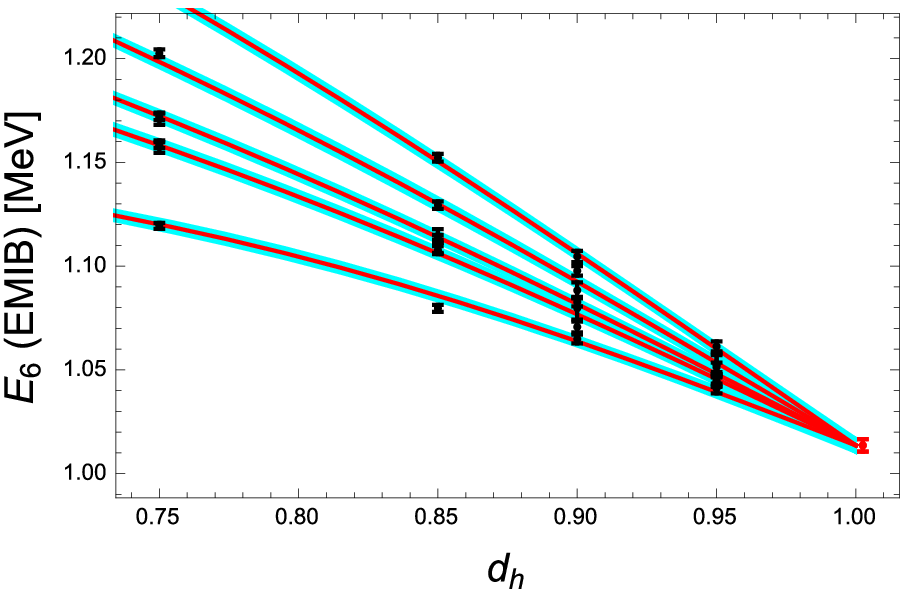}
\\ \vspace{.3cm}
\includegraphics[width=.45\linewidth]{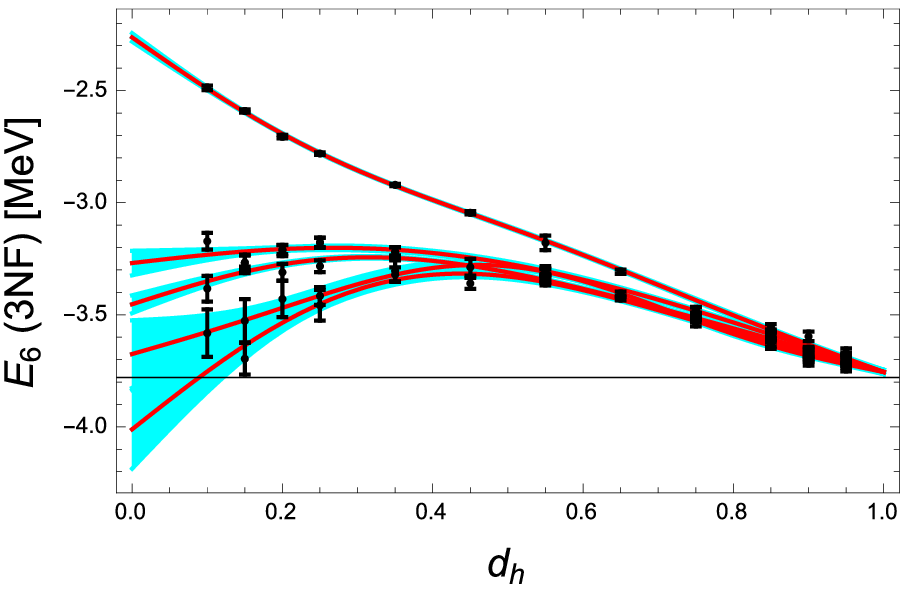}\quad\includegraphics[width=.45\linewidth]{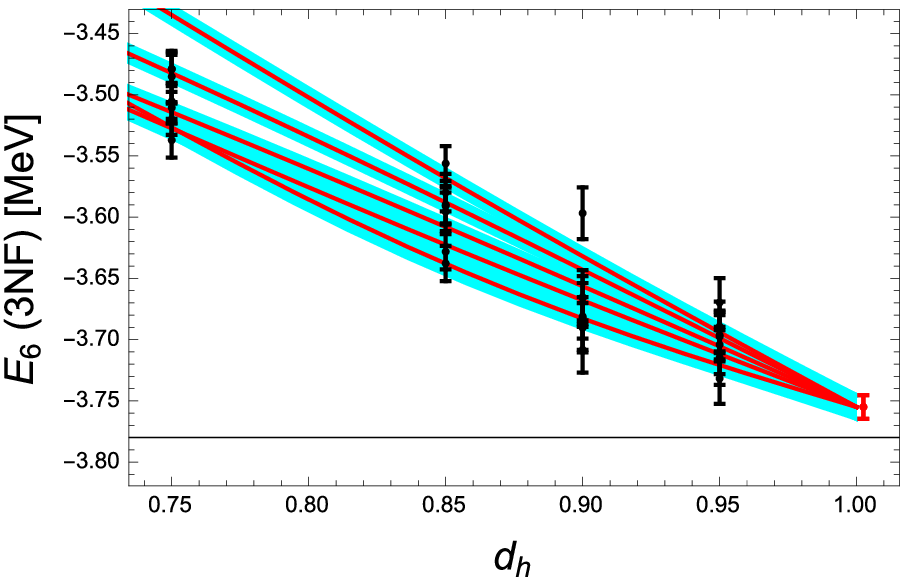}
\caption{\label{fig:He6 EMIB 3NF} PMC results for the electromagnetic and strong isospin-breaking (EMIB, upper panels) and NNLO three-nucleon force (3NF, lower panels) 
contributions to the ground state energy of $^6$He. The left panels show the full range between $d_h^{} = 0$ and $d_h^{} = 1$, while the right panels show a close-up near $d_h^{} = 1$. 
Each data point has been individually extrapolated to infinite Euclidean projection time. Notation and conventions are as for Fig.~\ref{fig:He6 LO NLO}.}
\end{figure}

The sign problem in the $A = 6$ system with 2~protons and 4~neutrons (or {\it vice versa}) is somewhat more severe than for $^{12}$C. Hence, if calculations are performed
entirely at $d_h^{} = 1$, the extrapolation to infinite Euclidean time (while still feasible) has to be performed using data with a rather limited range in $N_t^{}$. 
However, for $d_h^{} < 1$ this situation improves rapidly.
For $^6$He and $^6$Be, we shall therefore approach the problem differently than for $^{12}$C. We perform the extrapolation in Euclidean time for each pair of $C_4^{}$ and $d_h^{}$, with
the extrapolation $d_h^{} \to 1$ as the final step of the analysis, in terms of Eq.~(\ref{eqn:fit function}). For each value of $C_4^{}$ and $d_h^{}$, a constrained Euclidean time extrapolation is 
performed using a minimum of three trial states. This ``triangulation'' strategy is described in detail in Ref.~\cite{Lahde:2014sla}. 
Once the data have been extrapolated to infinite Euclidean time, the
resulting data are then subjected to the constrained global fitting procedure, as described for $^{12}$C. In Figs.~\ref{fig:He6 LO NLO} and~\ref{fig:He6 EMIB 3NF}, we show the fit results for the 
LO, NLO, EMIB, and 3NF contributions to the $^6$He energy. The explicit results are summarized in Table~\ref{tab:A6 results}. We find that the PMC results and the extrapolations for $A = 6$
are quantitatively similar to those of the $^{12}$C system. The extrapolations of our results to $d_h^{} = 1$ are in good agreement with direct calculations at $d_h^{} = 1$ (where the sign problem 
is maximal) and we also find no indication of a breakdown or inconsistencies in the Euclidean time extrapolations as $d_h^{} \to 1$.

\begin{table}[ht!]
\caption{\label{tab:A6 results} Contributions to the ground state energy of the $A = 6$ system after extrapolation $d_h^{} \to 1$. The contributions from the improved leading order 
amplitude~(LO), the two-nucleon force at next-to-leading order~(NLO), the electromagnetic and strong isospin breaking~(EMIB) and the three-nucleon force at next-to-next-to-leading order~(3NF)
are shown separately. The leftmost column shows the result of a direct calculation at $d_h^{} = 1$ without extrapolation in $d_h^{}$, and the other columns give the extrapolated results when
progressively more data is excluded in the vicinity of $d_h^{} = 1$.}
\vspace{.2cm}
\begin{tabular}{l|rrrrr}
& \multicolumn{1}{c}{$d_h^{} = 1$} & \multicolumn{1}{c}{$d_h^{} \leq 0.95$} & \multicolumn{1}{c}{$d_h^{} \leq 0.90$} & \multicolumn{1}{c}{$d_h^{} \leq 0.85$}
& \multicolumn{1}{c}{$d_h^{} \leq 0.75$} \\
\hline\hline
LO & $-27.49(7)$ & $-27.54(4)$ & $-27.56(6)$ & $-27.56(10)$ & $-27.34(46)$ \\
NLO & $2.61(4)$ & $2.58(2)$ & 2.61(3) & 2.66(4) & 3.05(11) \\
EMIB ($^6$He) & $1.021(6)$ & $1.014(3)$ &1.014(5) & 1.012(9) & 1.04(4) \\
EMIB ($^6$Be) & $2.65(1)$ & $2.66(1)$ &2.67(2) & 2.68(3)& 2.68(14) \\
3NF & $-3.77(3)$ & $-3.76(1)$ & -3.77(2)& -3.73(2) & -3.69(10)
\end{tabular}
\end{table}

\begin{figure}
\includegraphics[width=.45\linewidth]{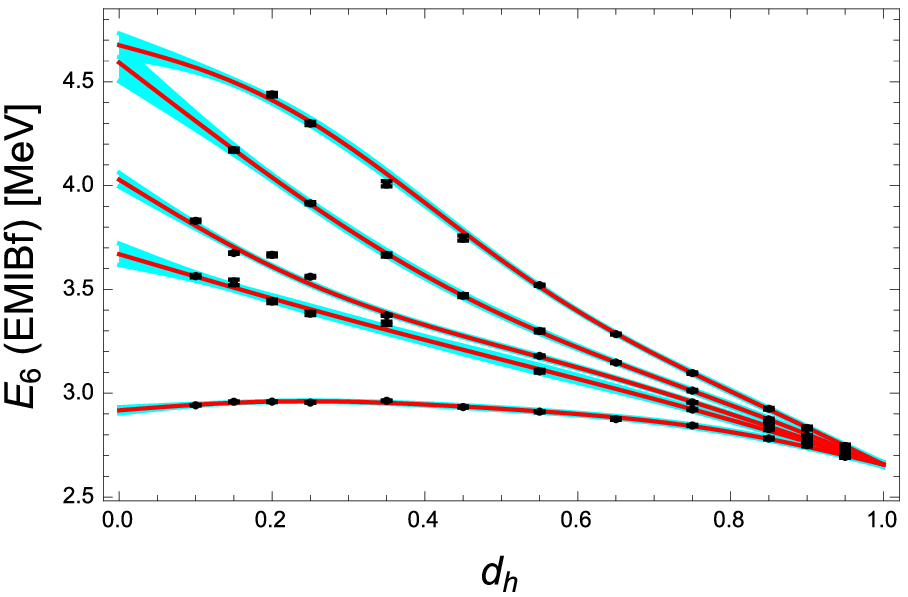}\quad\includegraphics[width=.45\linewidth]{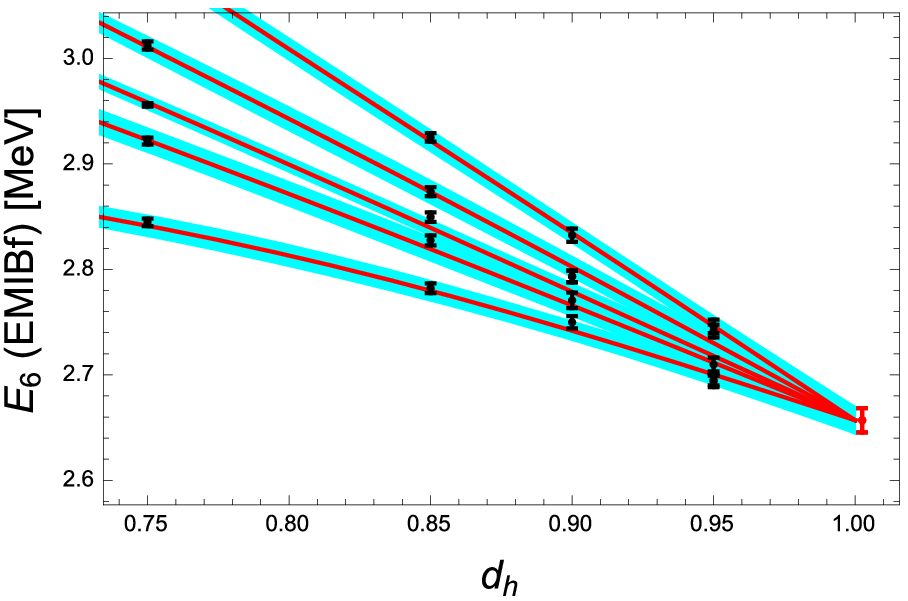}
\caption{\label{fig:Be6 EMIB} PMC results for the electromagnetic and strong isospin-breaking (EMIB) component of $^6$Be. The left panel shows the full range between 
$d_h^{} = 0$ and $d_h^{} = 1$, while the right panel shows a close-up near $d_h^{} = 1$. Each data point has been individually extrapolated to infinite Euclidean projection time.
Notation and conventions are as for Fig.~\ref{fig:He6 LO NLO}.}
\end{figure}

For the $A = 6$ system, we have performed PMC calculations for five values of $C_4^{}$, namely $-3.6 \times 10^{-5}$,
$-4.2 \times 10^{-5}$, $-4.4 \times 10^{-5}$,$-4.8 \times 10^{-5}$, and $-5.4 \times 10^{-5}$ (in units of~MeV$^{-2}$). 
We have employed a much larger range in $d_h^{}$, and moreover each PMC data point has an order of magnitude better
statistics compared with the data for $^{12}$C. In all figures related to $A = 6$, the uppermost band of data corresponds to $C_4^{} = -3.6 \times 10^{-5}$~MeV$^{-2}$
and the lowest one to $C_4^{} = -5.4 \times 10^{-5}$~MeV$^{-2}$. 
As the data at low $d_h^{}$ exhibit increased curvature, the order $n$ of our fit function has been taken to be $n = 3$ in contrast
to the $^{12}$C data, for which $n = 1$ was sufficient in most cases. For $^6$Be, the LO, NLO, and 3NF results are identical to those shown in Figs.~\ref{fig:He6 LO NLO} 
and~\ref{fig:He6 EMIB 3NF}. 
However, the electromagnetic part of the EMIB contribution changes due to the 
different numbers of protons and neutrons. In Fig.~\ref{fig:Be6 EMIB}, we show our results for the EMIB contribution to the energy of the $^6$Be ground state.

\begin{figure}
\includegraphics[width=.48\linewidth]{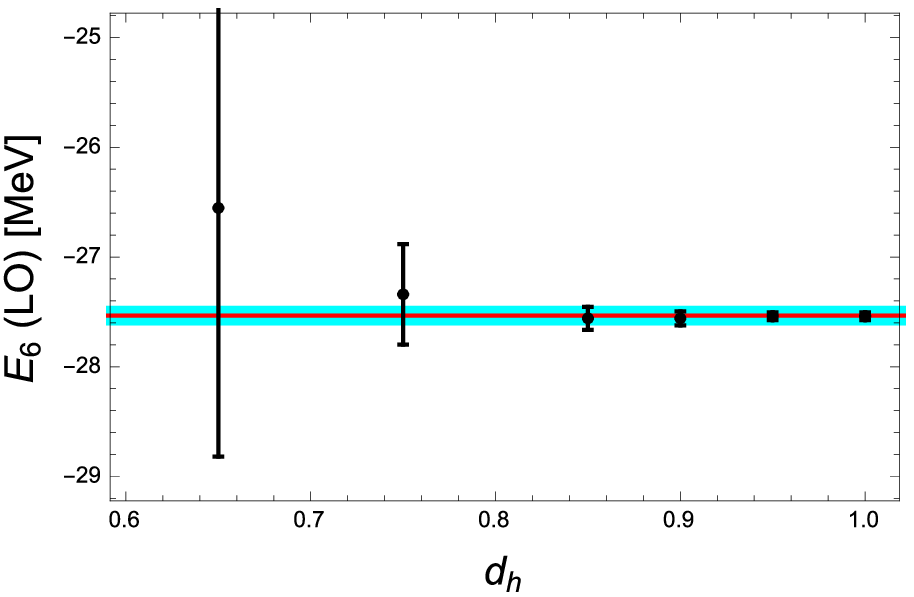} \\
\vspace{.02\linewidth}
\includegraphics[width=.48\linewidth]{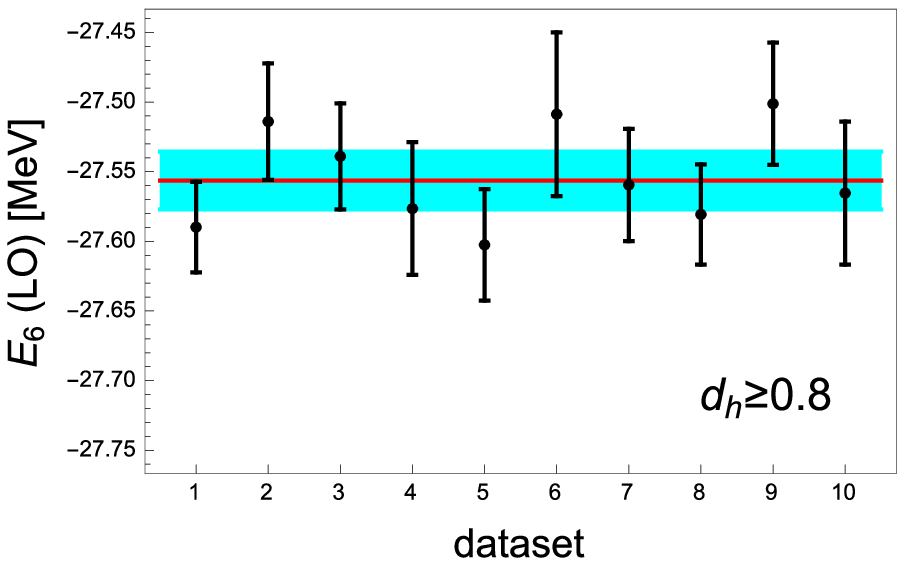} \hspace{.01\linewidth}
\includegraphics[width=.48\linewidth]{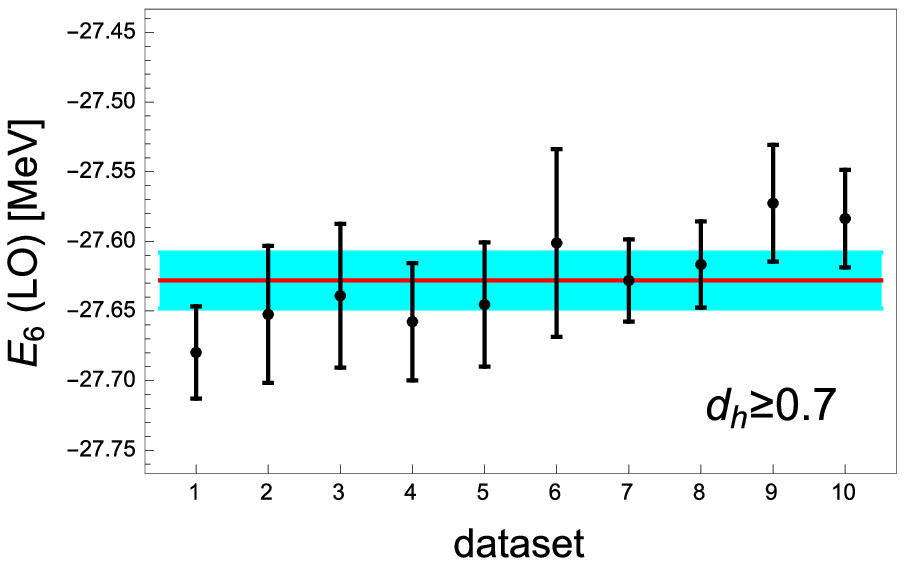}
\caption{\label{fig:He6 LO stability} Upper panel:
Stability of the $^6$He ground state energy at LO under extrapolation $d_h^{} \to 1$. The data point at $d_h^{} = 1$ represents a combined analysis including all the SSE data for
$d_h^{} < 1$ as well as the direct calculation at $d_h^{} = 1$ without SSE. The data point at $d_h^{} = 0.95$ only includes SSE data with $d_h^{} \leq 0.95$ {\it etc.} 
Lower panels: Ground state energy of $^6$He at LO for $d_h^{} \to1$, using SSE data with $d_h^{} \ge 0.8$ (left panel) and $d_h^{} \ge 0.7$ (right panel). These data are the result
of a linear extrapolation $d_h^{} \to 1$, such that each entry corresponds to one of ten possible datasets, as discussed in the main text. 
The horizontal band with 90\% confidence level is provided for visual guidance only.}
\end{figure}

As for $^{12}$C, we have also studied the stability of the extrapolated results for $A = 6$ as successive data points in the vicinity of $d_h^{} = 1$ are omitted. These results are 
summarized in Table~\ref{tab:A6 results}, and the behavior of the extrapolated LO energy is illustrated in Fig.~\ref{fig:He6 LO stability}. Our findings suggest that a satisfactory extrapolation
$d_h^{} \to 1$ can be obtained as long as PMC calculations can be performed for values no smaller than $d_h^{} \simeq 0.80$, at least for the present range of coupling constants
$C_4^{}$ employed in the analysis. We note that a larger range in $C_4^{}$ may allow smaller values of $d_h^{}$ to serve as a useful starting point for the extrapolation, especially if the linear
region around $d_h^{} = 1$ is thereby expanded. These findings appear consistent with our conclusions for $^{12}$C.

In contrast to performing extrapolations from SSE data below a given value of $d_h^{}$, we can instead perform \emph{linear} fits to SSE data \emph{above} a given $d_h^{}$. Such an analysis 
can demonstrate in what range of $d_h^{}$ a linear description remains valid in the vicinity of $d_h^{} = 1$. For this purpose, we perform linear fits for all possible combinations of three 
$C_4^{}$ subsets (of the five total). This provides ten different possible datasets. We then choose a value of $d_h^{}$ below which all data is discarded, and perform a linear extrapolation to
the remaining data points. In fig.~\ref{fig:He6 LO stability}, we show the extrapolated LO results for $d_h^{} \ge 0.7$ and $d_h^{} \ge 0.8$.  We also show the average of the fits and 
corresponding 90\% confidence bands.  For the analysis with $d_h^{} \ge 0.8$, our average is consistent with the complete LO extrapolation shown in Table~\ref{tab:A6 results}, while for the
analysis with $d_h^{} \ge 0.7$, a clear systematical error is found. This shows (at least for $^6$He) that a linear description is valid for $d_h^{} \ge 0.8$.

\begin{figure}
\includegraphics[width=.75\linewidth]{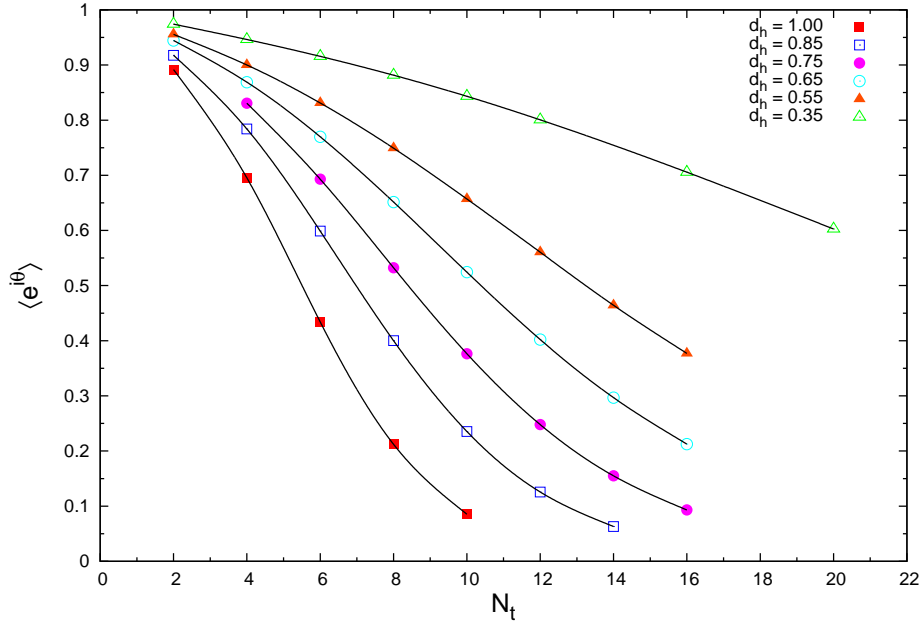}
\caption{\label{fig:lt vs sign} 
The mean value of the exponential of the complex phase $\langle e^{i\theta}\rangle$ of $\det({\bm M})$ as a function of Euclidean time step $N_t^{}$ and SSE parameter $d_h^{}$ for 
$C_4^{}=-4.8 \times 10^{-5}$~MeV$^{-2}$ in the $A=6$ system. The interpolating lines are intended as a guide to the eye.}
\end{figure}

As an explicit demonstration of the amelioration of the sign problem using the SSE method, we show in Fig.~\ref{fig:lt vs sign} for $A = 6$ the dependence of the mean value of the 
exponential of the complex phase $\langle e^{i\theta}\rangle$ of the PMC calculation on the number of Euclidean time steps $N_t^{}$ and the SSE parameter $d_h^{}$ 
for $C_4^{}=-4.8 \times 10^{-5}$~MeV$^{-2}$. For the case of $d_h^{}=1$, the mean value quickly approaches zero as $N_t^{}$ is increased, indicating that the sign problem is 
becoming severe at rather modest Euclidean projection times. For decreasing values of $d_h^{}$, the effect of the sign problem is successively diminished, allowing 
the PMC method to be extended to significantly larger values of $N_t^{}$.  

\section{Discussion \label{discussion}}

The SSE method introduced here is inspired by the existence of an SU(4) symmetric Hamiltonian which provides a reasonably accurate description of the
physics of the full Chiral EFT Hamiltonian. This has already proven useful in earlier work, as it greatly facilitates finding an accurate initial wave function which minimizes the extent of
Euclidean time projection necessary with the full Hamiltonian. We have here illustrated how this concept can be taken one step further, by studying a weighted sum of physical and
SU(4) symmetric Hamiltonians. In this way, the sign problem could be arbitrarily ameliorated, at the price of introducing an extrapolation in a control parameter $d_h^{} \to 1$. In practice, this
means that the SSE method is only useful as long as the extrapolation errors can be kept under control. Naturally, performing simulations at a range of values
of $d_h^{}$ has the potential to multiply the required CPU time by a large factor. However, we have found that we are able to avoid an exponential increase in computation time as a function
of Euclidean projection time, as long as we are able to perform simulations for $d_h^{} > 0.75$, as the accuracy of extrapolation then remains comparable with the statistical errors of 
typical simulations at $d_h^{} = 1$. We have also explored the freedom in the choice of the SU(4) symmetric Hamiltonian, which clearly plays no role at $d_h^{} = 1$, but which in general
gives different results for $d_h^{} \neq 1$. We have therefore made use of a ``triangulation'' method to improve the accuracy of the extrapolation $d_h^{} \to 1$.

An important consideration is whether a continuous shift from an SU(4) symmetric Hamiltonian to the full Chiral EFT Hamiltonian can be effected without inducing 
non-trivial changes in the spectrum. For instance, the appearance of a level crossing at a critical value of $d_h^{}$
would clearly limit the applicability of SSE. We note that such level crossings as a function of $d_h^{}$ would be quite rare for low-lying nuclear bound states, and furthermore we 
have the freedom to choose $C_4^{}$ to avoid such level crossings. But if a level crossing were to occur, there would be some subtleties in obtaining accurate and converged results.  
For such cases, it would be preferable to take the $d_h^{} \to 1$ limit first, followed by extrapolation in Euclidean time. An even better solution would entail solving a coupled-channel problem
using multiple initial states. This makes it possible to 
disentangle one or more nearly degenerate states with the same quantum numbers. So far, we have concentrated on systems where such degeneracies are 
not expected, and where PMC is still possible (though difficult) without the SSE method. We have also not yet explored different "extrapolation Hamiltonians", where the sign oscillations are 
minimized while retaining as much as possible of the full Chiral EFT structure.

In this first report on the SSE method, 
we have presented extensive results for $^6$He and $^{12}$C, for a wide range of $d_h^{}$ and SU(4) symmetric Hamiltonians. For the case of $^6$He, we have first extrapolated all
results to infinite Euclidean time, followed by an extrapolation $d_h^{} \to 1$. For the case of $^{12}$C, we have performed the extrapolation $d_h^{} \to 1$ directly for data at finite Euclidean
projection time. In both cases, we find similar behavior as a function of $d_h^{}$ and excellent linearity in the vicinity of $d_h^{} = 1$. Which ordering of the limits is preferable depends on
how severe the sign problem is for a given system at $d_h^{} = 1$. In cases where the sign problem is severe, the method of first taking the limit $d_h^{} \to 1$ followed by an extrapolation
in Euclidean time is preferable. This is because the Euclidean time extrapolation may be ambiguous if only very short Euclidean times are accessible with PMC.
Our results provide the first determination of the ground state energy of $^6$He using lattice Chiral EFT, 
and for $^{12}$C we find that our new results at larger Euclidean projection time agree very well with previous infinite-time extrapolations. 

We have presented here the first studies of the binding energy of $^6$He within lattice Chiral EFT. While the results are very encouraging for the feasibility of calculations for neutron-rich
systems, we also find that $^6$He appears underbound by $\simeq 1$~MeV at NNLO. Nevertheless, since the present
calculations are performed in an $L = 6$ box, the possibility remains that finite volume effects could significantly improve on the current situation. In particular, since $P$-wave states are
underbound in a finite volume~\cite{Konig:2011nz,Konig:2011ti} while $S$-wave states are overbound, calculations in larger boxes (or extrapolations to $L = \infty$) would shed more
light on this situation.

We acknowledge partial financial support from the 
Deutsche Forschungsgemeinschaft (Sino-German CRC 110), the Helmholtz Association (Contract No.\ VH-VI-417), 
BMBF (Grant No.\ 05P12PDFTE), and the U.S. Department of Energy (DE-FG02-03ER41260). Further support
was provided by the EU HadronPhysics3 project and the ERC Project No.\ 259218 NUCLEAREFT. The computational resources 
were provided by the J\"{u}lich Supercomputing Centre at  Forschungszentrum J\"{u}lich and by RWTH Aachen. TL acknowledges financial support
from the Magnus Ehrnrooth Foundation of the Finnish Society of Sciences and Letters.

\end{document}